\documentclass{optica-article}

\journal{opticajournal} 

\articletype{Research Article}
\usepackage{booktabs}
\usepackage{caption}
\usepackage{subcaption}
\usepackage{rotating} 
\usepackage{silence}
\usepackage{hyperref}
\hypersetup{
colorlinks=true,
citecolor=blue,
}

\usepackage{makecell}
\usepackage{amsmath}
\usepackage{siunitx}
\allowdisplaybreaks[4]
\AtBeginDocument{\hfuzz=100pt \vfuzz=100pt \hbadness=10000 \vbadness=10000}
\begin{document}

\title{Analytical Modeling of Far-Field Wavefront Error with Beam-Waist and Lateral-Shift Effects in Spaceborne Laser Interferometry}
\author{Ya-Zheng Tao,\authormark{1,2}$^\dagger$ Rui-Hong Gao,\authormark{3}$\dagger$ Guangzhou Xu,\authormark{1,2} and Yue-Liang Wu,\authormark{2,4,5}$^\ast$}

\address{
\authormark{1}School of Fundamental Physics and Mathematical Sciences, Hangzhou Institute for Advanced Study, UCAS, Hangzhou 310024, Zhejiang Province, China\\
\authormark{2}University of Chinese Academy of Sciences, Beijing 100049, China\\
\authormark{3}Center for Gravitational Wave Experiment, National Microgravity Laboratory, Institute of Mechanics,
Chinese Academy of Sciences, Beijing 100190, China\\
\authormark{4}The International Centre for Theoretical Physics Asia-Paciﬁc, University of Chinese Academy of Sciences, Beijing 100190, China\\
\authormark{5}CAS Key Laboratory of Theoretical Physics, Institute of Theoretical Physics,Chinese Academy of Sciences, Beijing 100190, China
}
\noindent\authormark{$\dagger$} These two authors contributed equally.\\
\noindent\authormark{$\ast$} Corresponding author\\
\email{\authormark{$\dagger$}taoyazheng@ucas.ac.cn}
\email{\authormark{$\dagger$}gaoruihong@imech.ac.cn}
\email{\authormark{$\ast$}ylwu@itp.ac.cn }


\begin{abstract*}
	The coupling between far-field wavefront error (WFE) and laser pointing jitter is an important source of tilt-to-length (TTL) noise in spaceborne laser interferometric links. We extend the Nijboer--Zernike analytical model for far-field WFE of truncated Gaussian beams by incorporating two practical initial-condition parameters, the beam-waist-to-aperture ratio $q$ and the normalized lateral spot-shift ratio $s_r$, to account for realistic beam truncation and alignment conditions.
	Based on this model, we analyze the influence of $q$ on far-field WFE in addition to the conventional received-power trade-off, showing that decreasing $q$ from 1 to 0.9 and from 0.9 to 0.8 reduces the mean far-field WFE by approximately 10\% and 14\%, respectively, in Monte Carlo simulations of random initial aberrations. We also derive the direct contribution of lateral spot shift and its coupling with transmitted WFE (constrained to $\lambda/20$). For the normalized lateral spot-shift ratio $s_r$, a $2~\mu\mathrm{m}$ entrance-pupil displacement in a Taiji-like telescope corresponds to $s_r=0.001$ and produces a phase-angle coupling coefficient of about $0.0892~\mathrm{pm/nrad}$, close to the typical far-field TTL requirement $0.1~\mathrm{pm/nrad}$, while the spot-shift--aberration coupling terms are much smaller and can be neglected in practical tolerance estimation.	These results provide a theoretical basis for beam-parameter optimization and alignment tolerance design in future space-based gravitational-wave detection missions.
	\end{abstract*}

\section{Introduction}\label{se:1}
The direct detection of gravitational waves has opened a new observational window for astronomy and fundamental physics, and space-based gravitational wave detection is expected to further extend this window to the low-to-mid-frequency band ($0.1~\mathrm{mHz}$ to $1~\mathrm{Hz}$). Representative missions, such as LISA, Taiji, and TianQin, are designed to employ long-baseline laser interferometry between drag-free spacecraft arranged in triangular constellations \cite{jennrich2009lisa,luo2020brief,luo2016tianqin}. In these missions, gravitational wave signals are encoded in extremely small variations in the inter-spacecraft interferometric links. Therefore, achieving pm/$\sqrt{\text{Hz}}$-level displacement sensitivity requires not only precise interferometric readout but also careful suppression of various noise sources within the measurement band.

Among displacement noise sources, tilt-to-length (TTL) coupling noise is considered one of the dominant secondary noise sources after laser frequency noise \cite{wanner2024depth}. The coupling between far-field wavefront error (WFE) and laser pointing jitter is a typical manifestation of the non-geometric TTL effect. Ideally, an undistorted transmitted beam forms an approximately spherical wavefront after long-distance propagation, so small pointing variations do not induce additional phase fluctuations in the received beam. In practice, however, telescope aberrations, alignment errors, thermal deformation, and other imperfections distort the transmitted wavefront and generate far-field WFE. The resulting far-field WFE further couples with laser pointing jitter, leading to a pointing-angle-dependent phase offset in the received top-hat beam \cite{robertson1997optics}. Since million-kilometer-scale propagation cannot be fully reproduced in ground experiments, this coupling noise is usually investigated through theoretical modeling and numerical simulation.

Previous studies on far-field WFE and pointing-jitter coupling have generally followed three methodological routes: approximate expansions of scalar diffraction formulas, numerical computation, and analytical approximations based on Nijboer-Zernike theory. The first route involves approximate expansions of scalar diffraction formulas. Bender analyzed this problem using the Fresnel-Kirchhoff diffraction integral, expanded the exit-pupil function into a third-order power series, and evaluated the far-field coupling between low-order aberrations and pointing jitter \cite{bender2005wavefront}. Sasso et al. further investigated the coupling between low-order Zernike aberrations and tip/tilt by examining the far-field phase under different transmitted-beam tilt conditions, and used Monte Carlo simulations to assess the effect of random wavefront distortions \cite{sasso2018coupling,sasso2018coupling2}. Zhao et al. extended this framework to higher-order telescope aberrations and prototype telescope systems \cite{zhao2020tilt,zhao2021far}. Xiao et al. studied higher-order aberrations and clarified the role of stationary points in suppressing coupling noise \cite{xiao2023analysis}. Chen et al. applied this type of analysis to telescope optimization by comparing the contributions of different aberrations to coupling noise \cite{chen2022reducing}.

The second route relies on numerical computation and optical simulation. Waluschka performed an early simulation-based evaluation of the far-field wavefront over a LISA-like arm length \cite{waluschka1999lisa}. Kenny later derived a semi-analytical formulation based on Nijboer-Zernike theory and evaluated the resulting diffraction integrals numerically \cite{kenny2020beam}. Weaver et al. and Tao et al. modeled the propagation of distorted beams and estimated the corresponding far-field WFE and pointing-jitter coupling using the Mode Expansion Method (MEM) and Gaussian Beamlet Decomposition (GBD), respectively \cite{weaver2020analytic,tao2023estimation}.

The third route is based on analytical approximations derived from Nijboer-Zernike theory. Vinet et al. introduced this approach by applying a first-order expansion of the transmitted WFE and deriving an analytical expression for the influence of individual aberrations on pointing-jitter coupling \cite{vinet2020numerical}. Tao et al. further extended this framework to higher-order aberration expansions, revealing the mechanisms by which individual aberrations and their higher-order coupling terms contribute differently to far-field WFE \cite{tao2025approximate}. Fan et al. applied Nijboer-Zernike theory to Fresnel diffraction under a first-order expansion of the transmitted WFE, enabling the description of truncated Gaussian beams whose waist position does not coincide with the entrance pupil \cite{fan2026nijboer}.

Compared with approximate expansions of scalar diffraction formulas, the Nijboer-Zernike-based approach provides a more rigorous mathematical framework and clearer physical interpretability. It is also more suitable for establishing accurate parameter mappings between the laser link and other subsystems in system-level simulations. Compared with purely numerical methods, it offers much higher computational efficiency while maintaining sufficient accuracy, although at the cost of more involved theoretical derivations. In this study, we extend the far-field WFE model established in previous work by introducing two additional initial-condition parameters: the beam-waist-to-aperture ratio $q$ and the normalized lateral spot-shift ratio $s_r$. The parameter $q$ describes cases in which the waist radius of the emitted Gaussian beam differs from the telescope aperture radius, while $s_r$ characterizes the radial displacement between the beam spot center and the telescope aperture center. These two parameters allow the model to better represent practical experimental designs and realistic assembly conditions. Specifically, Section \ref{se:2} introduces the theoretical model of far-field WFE based on Nijboer-Zernike theory. Section \ref{se:3} discusses the trade-off associated with the beam-waist-to-aperture ratio $q$ in terms of the received laser power at the remote spacecraft, and analyzes how variations in $q$ affect both the received power and far-field WFE, thereby providing a strategy for selecting $q$. Section \ref{se:4} investigates the influence of the normalized lateral spot-shift ratio $s_r$ on far-field WFE, including the coupling between the initial beam spot offset and the initial WFE. Based on the scientific measurement requirements, a quantitative constraint on the allowable spot offset of the incident beam at the telescope pupil is then derived. Finally, Section \ref{se:5} summarizes the conclusions.

\section{Nijboer-Zernike Theory and the expression of far-field wavefront error} \label{se:2}
Excluding the wavefront errors already present in the laser before it enters the telescope, the wavefront errors induced by the telescope mainly arise from optical path differences among different regions within the telescope, which are independent of the amplitude distribution. Accordingly, a lateral offset of the incident beam spot center at the exit pupil may arise from assembly errors, beam-centering deviations, or other alignment-related factors, but it does not directly modify the phase distribution induced by the telescope. We can therefore assume that the electric field at the exit pupil is

\begin{equation}\label{wfegeneraldefinition}
	E_a({r_{t}}; z=0)=E^s_0({r_{t}}; z=0)e^{i\varOmega_a},	
\end{equation}
where $E^s_0({r_{t}}; 0)$ is the electric field of a shifted truncated circular Gaussian beam:
\begin{equation}
	E^s_0({r_{t}}; 0)=
	\left\{
	\begin{aligned}
		\sqrt{\frac{2P_0}{\pi{w_0^2}}}e^{-\left[{(r_{t}\cos\theta-s_x)}^2+{(r_{t}\sin\theta-s_y)}^2\right]/w_0^2},&&\mbox{if ${r_{t}}\leq{r_a}$ } \\
		0,&&\mbox{if ${r_{t}}>{r_a}$ } 
	\end{aligned}
	\right.
\end{equation}
where $w_0$ is the waist radius of the Gaussian beam and $r_a$ is the telescope aperture radius. $P_0$ is the laser power. $s_x$ and $s_y$ represent the lateral spot shifts in the x and y directions, respectively. $\varOmega_a$ in \eqref{wfegeneraldefinition} is the transmitted WFE, which is described by a combination of Zernike polynomials:
\begin{equation}\label{AberrationDefinition}
	\varOmega_a(\rho,\theta)=\sum_{n=0}^{N}\sum_{m=-n}^{n} a^m_n\,Z^m_n(\rho,\theta),
\end{equation}
where $\rho={r_{t}}/r_a$ is restricted to the unit disk ($0\leq\rho\leq1$), and $\theta$ is the azimuthal angle. The condition $n-m\geq0$ is satisfied, and $n-m$ is even. $a^m_n$ denotes the corresponding coefficient, and $Z^m_n$ denotes the Zernike polynomial written using OSA/ANSI indexing. $N$ represents the maximum value of $n$. The Zernike polynomials are defined on the unit disk as
 \begin{equation}
	Z^m_n(\rho,\theta)=
	\left\{
	\begin{aligned}
		R^m_n(\rho)\cos\left(m\theta\right),&&\mbox{if $m\geq0$ } \\
		R^{-m}_n(\rho)\sin\left(-m\theta\right),&&\mbox{if $m<0$ }
	\end{aligned}
	\right.
\end{equation}
where $R^m_n(\rho)$ is the radial polynomial:
\begin{equation}
	R_n^{|m|}(\rho)=(-1)^{(n-|m|)/2}\rho^{|m|}P_{(n-|m|)/2}^{(|m|,0)}(1-2\rho^2) ,
\end{equation}
where $P_{t}^{(\alpha,\beta)}$ is the Jacobi polynomial of degree $t$, and $R^m_n(\rho)$ satisfies the orthogonality relation:
\begin{equation}
	\int_0^1R_n^{|m|}(\rho)R_{n'}^{|m|}(\rho)\rho{d}\rho=\frac{\delta_{n,n'}R_n^{|m|}(1)}{2(n+1)}.
\end{equation}

\begin{figure}
	\centering
	\begin{subfigure}[b]{0.45\textwidth}
		\centering
		\caption{Expansion of $e^{-\rho^2/{q^2}}$  at l=2}
		\includegraphics[width=1\textwidth]{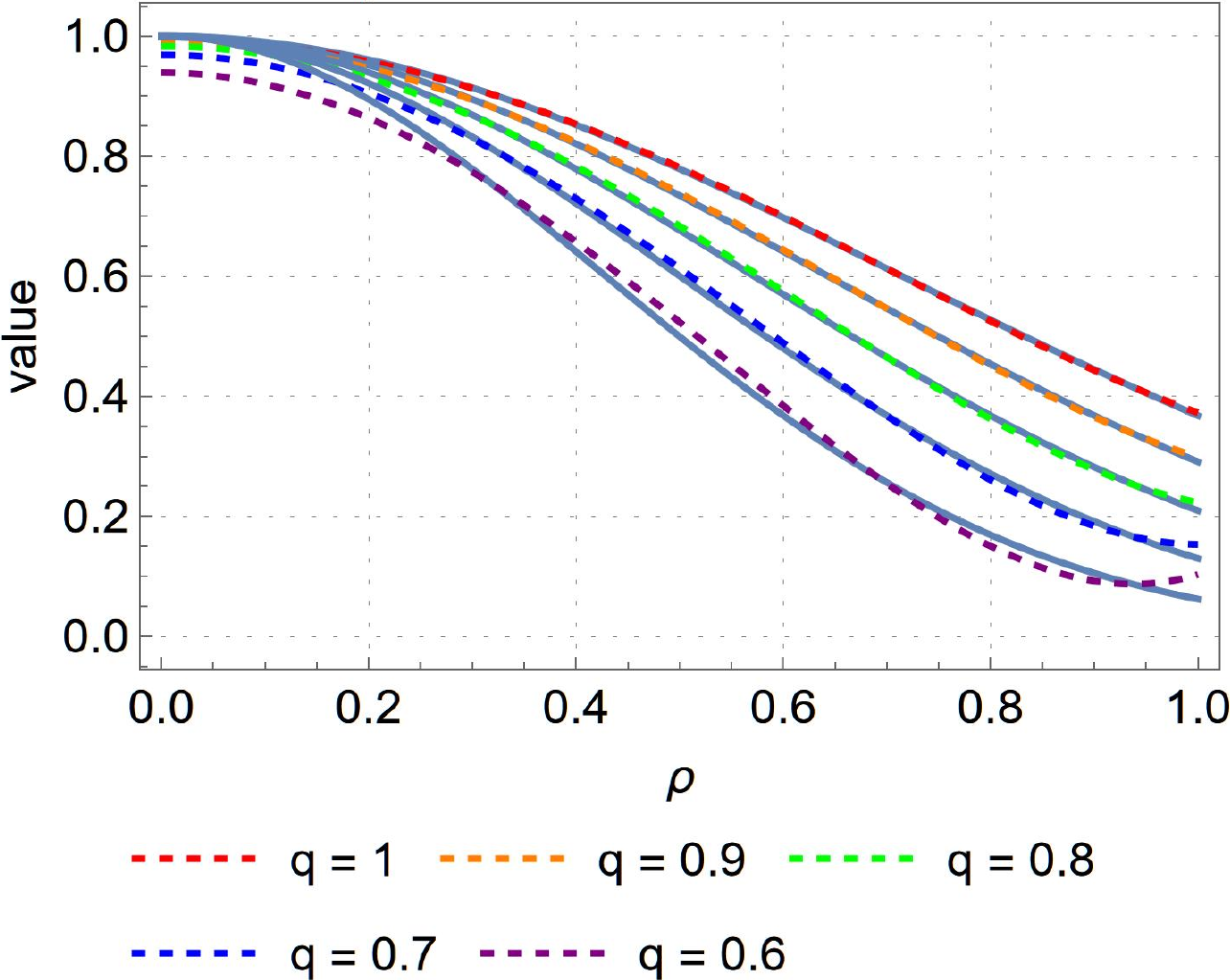}
		\label{fig:l=2}
	\end{subfigure}
	\begin{subfigure}[b]{0.45\textwidth}
		\centering
		\caption{Expansion of $e^{-\rho^2/{q^2}}$  at l=3}
		\includegraphics[width=1\textwidth]{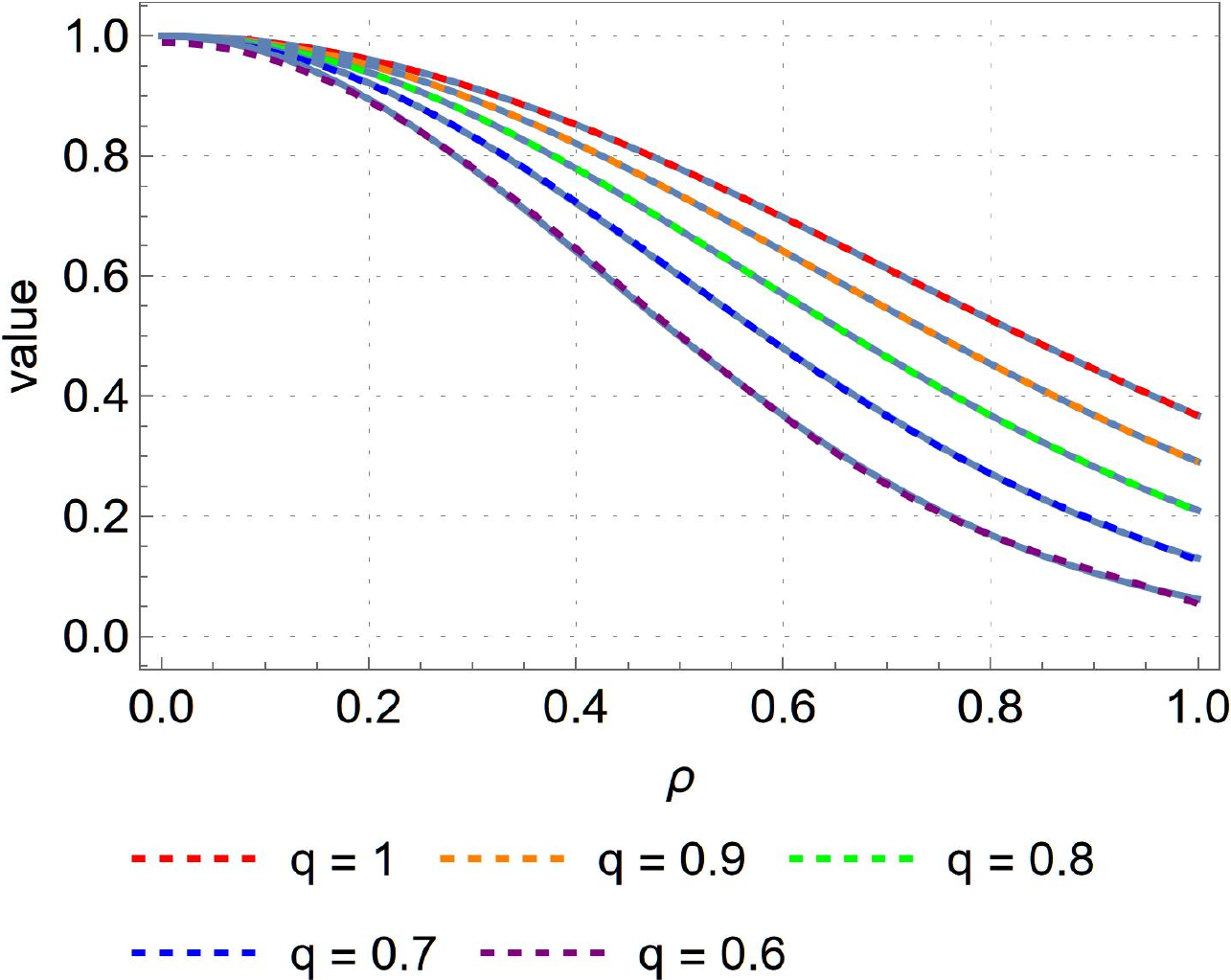}
		\label{fig:l=3}
	\end{subfigure}    
	   \caption{Accuracy comparison for the expansion of $e^{-\rho^2/q^2}$ truncated at $l=2$ (left) and $l=3$ (right).}
	   \label{fig:l_difference}
\end{figure}

Because the beam pointing angle is negligible compared with the propagation distance, the diffraction integral satisfies the paraxial approximation. Consequently, the far-field electric field can be expressed using Fraunhofer diffraction:
\begin{equation}\label{FraunhoferAberration1}
	E(r, \psi, z)=\sqrt{\frac{2P_0}{\pi{w_0^2}}}\frac{e^{ikz}e^{\frac{ik}{2z}r^2}}{i\lambda{z}}{r_a}^2{e^{-s_r^2}}
	\int_0^1\int_0^{2\pi}e^{-\rho^2/q^2}e^{2{s_r}\rho/q\cos(\theta-{\theta}_0)}e^{i\varOmega_a(\rho,\theta)}e^{-iv\rho\cos{(\theta-\psi)}}\rho{d}\rho{d}\theta.
\end{equation}
Here, we define three factors: the beam-waist-to-aperture ratio $q=w_0/r_a$, the normalized lateral spot-shift ratio $s_r={\sqrt{s_x^2+s_y^2}}/w_0$, and $v=\frac{k}{{z}}{r_a}r$. In addition, $\cos{\theta}_0=s_x/{\sqrt{s_x^2+s_y^2}}$ and $\sin{\theta}_0=s_y/{\sqrt{s_x^2+s_y^2}}$. The integral part in \eqref{FraunhoferAberration1} is given by
\begin{equation}\label{FraunhoferAberration2}
	U(r, \psi, z)=\int_0^1\int_0^{2\pi}e^{-\rho^2/q^2}e^{2{s_r}\rho/q\cos(\theta-{\theta}_0)}e^{i\varOmega_a(\rho,\theta)}e^{-iv\rho\cos{(\theta-\psi)}}\rho{d}\rho{d}\theta.
\end{equation}
For $e^{-\rho^2/q^2}$ inside the integral, we apply the Bauer formula \cite{watson1922treatise}:
\begin{equation}\label{Bauer1}
	e^{-z\cos\phi}=(\frac{\pi}{2z})^{\frac{1}{2}}\sum_{l=0}^{\infty}{(-1)^l(2l+1)I_{l+\frac{1}{2}}(z)}{P_l(\cos{\phi})},	
\end{equation}
where $I_{l+\frac{1}{2}}$ is the modified Bessel function of the first kind, and $P_l$ is the Legendre polynomial. By setting $\cos\phi=2{\rho}^2-1$ and $z=\frac{1}{2q^2}$, and using the relation $P_l(2{\rho}^2-1)=R_{2l}^0(\rho)$, we obtain
\begin{equation}\label{Bauer2}
e^{-\rho^2/q^2}=e^{-\frac{1}{2q^2}}e^{-\frac{1}{2q^2}(2{\rho}^2-1)}=qe^{-\frac{1}{2q^2}}{\pi}^{\frac{1}{2}}\sum_{l=0}^{\infty}{(-1)^l(2l+1)I_{l+\frac{1}{2}}(\frac{1}{2q^2})}{R_{2l}^{0}(\rho)}.
\end{equation}
For the parameter range considered in this study ($0.6\leq q\leq 1$), truncating the expansion at $l=3$ provides sufficient accuracy, as shown in Fig.~\ref{fig:l_difference}.

For $e^{2{s_r}\rho/q\cos(\theta-{\theta}_0)}$, comparison of the numerical integration results obtained with different expansion orders shows that a first-order expansion is sufficient. For $e^{i\varOmega_a(\rho,\theta)}$, a second-order Taylor expansion is sufficient \cite{tao2025approximate}. Therefore, \eqref{FraunhoferAberration2} can be expanded into four terms:
\begin{subequations}\label{U_Expand}
\begin{align}
&U(r, \psi, z)=U_0(r, \psi, z)+U_1(r, \psi, z)+U_2(r, \psi, z)+U_3(r, \psi, z),\\
 \label{U_0}
&U_0(r, \psi, z)=\int_0^1\int_0^{2\pi}e^{-\rho^2/q^2}e^{-iv\rho\cos{(\theta-\psi)}}\rho{d}\rho{d}\theta,\\
 \label{U_1}
&U_1(r, \psi, z)=\int_0^1\int_0^{2\pi}e^{-\rho^2/q^2}(i\varOmega_a-\frac{1}{2}{\varOmega_a^2})e^{-iv\rho\cos{(\theta-\psi)}}\rho{d}\rho{d}\theta,\\
 \label{U_2}
&U_2(r, \psi, z)=2{s_r}/q\int_0^1\int_0^{2\pi}e^{-\rho^2/q^2}\cdot\rho\cos(\theta-{\theta}_0)\cdot{e^{-iv\rho\cos{(\theta-\psi)}}}\rho{d}\rho{d}\theta,\\
\label{U_3}
 &U_3(r, \psi, z)=2{s_r}/q\int_0^1\int_0^{2\pi}e^{-\rho^2/q^2}\cdot\rho\cos(\theta-{\theta}_0)\cdot(i\varOmega_a-\frac{1}{2}{\varOmega_a^2})\cdot{e^{-iv\rho\cos{(\theta-\psi)}}}\rho{d}\rho{d}\theta. 
\end{align}
\end{subequations}

The first two terms, $U_0(r, \psi, z)$ and $U_1(r, \psi, z)$, are referred to as the base term and the distortion term, respectively. Their derivation and discussion can be found in previous work \cite{tao2025approximate}. Previous work also indicates that the influence of all aberration terms on the far-field wavefront error can be expressed as a superposition of Bessel functions, with each term contributing to either the real or imaginary part. Moreover, retaining Bessel orders up to the fourth order is sufficient to meet the required numerical accuracy. Taking $U_0(r, \psi, z)$ as an example, its expression is given by
\begin{equation}\label{U0_expanded}
\begin{aligned}
U_0(r, \psi, z)&=qe^{-\frac{1}{2q^2}}{\pi}^{\frac{1}{2}}(2\pi)\left\{I_{\frac{1}{2}}\frac{J_1(v)}{v}+3I_{\frac{3}{2}}\frac{J_3(v)}{v}\right\}\\
&=qe^{-\frac{1}{2q^2}}{\pi}^{\frac{1}{2}}(2\pi)\left\{{\sigma}_0\frac{J_1(v)}{v}+{\tau}_0\frac{J_3(v)}{v}\right\},
\end{aligned}
\end{equation}
where
\begin{gather*}\label{co00}
	\left({\sigma}_0,\;{\tau}_0\right)=\left(I_{\frac{1}{2}},\;3I_{\frac{3}{2}}\right).
\end{gather*}
Here, $I_{l+\frac{1}{2}}$ denotes $I_{l+\frac{1}{2}}(\frac{1}{2q^2})$. In Appendix~\ref{se:6}, we provide the expressions and coefficients of the first 21 aberration terms, and their coupling terms associated with the $q$-correction. The first 21 Zernike polynomials are listed in Table~\ref{ZernikeList}. The influence of $q$ on the received laser power at the remote spacecraft and on the far-field wavefront error is discussed in Section~\ref{se:3}.
\begin{table*}
	\abovetopsep=0pt
	\aboverulesep=0pt
	\belowrulesep=0pt
	\belowbottomsep=0pt
		\begin{center}
				\caption{\textbf{The first 21 Zernike polynomials}}
			\begin{tabular}{c|c|c|c}
				\toprule[1.5pt]
				\textbf{Order} & \textbf{Aberration} & \textbf{Term} &  \textbf{Value} \\ 
				\midrule[1.5pt]
				1 & X-Tilt  & $Z_1^{1}$ & $\rho\cos{\phi}$\\
				\hline
				2 & Y-Tilt  & $Z_1^{-1}$ & $\rho\sin{\phi}$\\
				\hline
				3 & Defocus  & $Z_2^{0}$ & $2{\rho}^2-1$\\
				\hline	
				4 & $0^{\circ}$ Astigmatism  & $Z_2^{2}$ & ${\rho}^2\cos{2\phi}$\\
				\hline
				5 & $45^{\circ}$ Astigmatism  & $Z_2^{-2}$ & ${\rho}^2\sin{2\phi}$\\
				\hline
				6 & X-Coma  & $Z_3^{1}$ & $(3{\rho}^3-2\rho)\cos{\phi}$\\
				\hline
				7 & Y-Coma  & $Z_3^{-1}$ & $(3{\rho}^3-2\rho)\sin{\phi}$\\
				\hline
				8 & X-Trefoil  & $Z_3^{3}$ & ${\rho}^3\cos{3\phi}$\\
				\hline
				9 & Y-Trefoil  & $Z_3^{-3}$ & ${\rho}^3\sin{3\phi}$\\
				\hline
				10 & Spherical  & $Z_4^{0}$ & $6{\rho}^4-6{\rho}^2+1$\\
				\hline
				11 & X 2nd Astigmatism  & $Z_4^{2}$ & $(4{\rho}^4-3{\rho}^2)\cos{2\phi}$\\
				\hline	
				12 & Y 2nd Astigmatism  & $Z_4^{-2}$ & $(4{\rho}^4-3{\rho}^2)\sin{2\phi}$\\
				\hline
				13 & X-Tetrafoil  & $Z_4^{4}$ & ${\rho}^4\cos{4\phi}$\\
				\hline
				14 & Y-Tetrafoil  & $Z_4^{-4}$ & ${\rho}^4\sin{4\phi}$\\
				\hline
				15 & X 2nd Coma  & $Z_5^{1}$ & $(10{\rho}^5-12{\rho}^3+3{\rho})\cos{\phi}$\\
				\hline
				16 & Y 2nd Coma  & $Z_5^{-1}$ & $(10{\rho}^5-12{\rho}^3+3{\rho})\sin{\phi}$\\
				\hline
				17 & X 2nd Trefoil  & $Z_5^{3}$ & $(5{\rho}^5-4{\rho}^3)\cos{3\phi}$\\
				\hline
				18 & Y 2nd Trefoil  & $Z_5^{-3}$ & $(5{\rho}^5-4{\rho}^3)\sin{3\phi}$\\
				\hline
				19 & X-Pentafoil  & $Z_5^{5}$ & ${\rho}^5\cos{5\phi}$\\
				\hline
				20 & Y-Pentafoil  & $Z_5^{-5}$ & ${\rho}^5\sin{5\phi}$\\
				\hline
				21 & 2nd Spherical  & $Z_6^{0}$ & $20{\rho}^6-30{\rho}^4+12{\rho}^2-1$\\
				\bottomrule[1.5pt]	
			\end{tabular}
				\label{ZernikeList}
		\end{center}
\end{table*}

The third term $U_2(r, \psi, z)$ represents the direct influence of the spot shift itself on the far-field wavefront error, and we refer to it as the spot-shift term. The fourth term $U_3(r, \psi, z)$ accounts for the influence of the coupling between the spot shift and the transmitted wavefront error on the far-field wavefront error, and we refer to it as the spot-shift--distortion coupling term. The effects of these two terms are discussed in Section~\ref{se:4}.

\section{Beam-waist-to-aperture ratio \textit{q}}\label{se:3}

\subsection{The clipping trade-off in received power} \label{sbse:3.1}

The choice of the beam-waist-to-aperture ratio $q=w_0/r_a$ is governed by a well-known clipping trade-off in the optical power received at the remote aperture \cite{vinet2019lisa}. For a fixed laser power $P_0$, increasing $q$ increases the power loss caused by clipping at the transmitting aperture. Meanwhile, a larger $q$ reduces the far-field divergence angle and improves the energy concentration near the beam center after long-distance propagation. These competing effects give rise to an optimal value of $q$ that maximizes the optical power collected by the remote receiving aperture.

\begin{figure}[htbp]
	\centering
	  \includegraphics[width=0.6\textwidth]{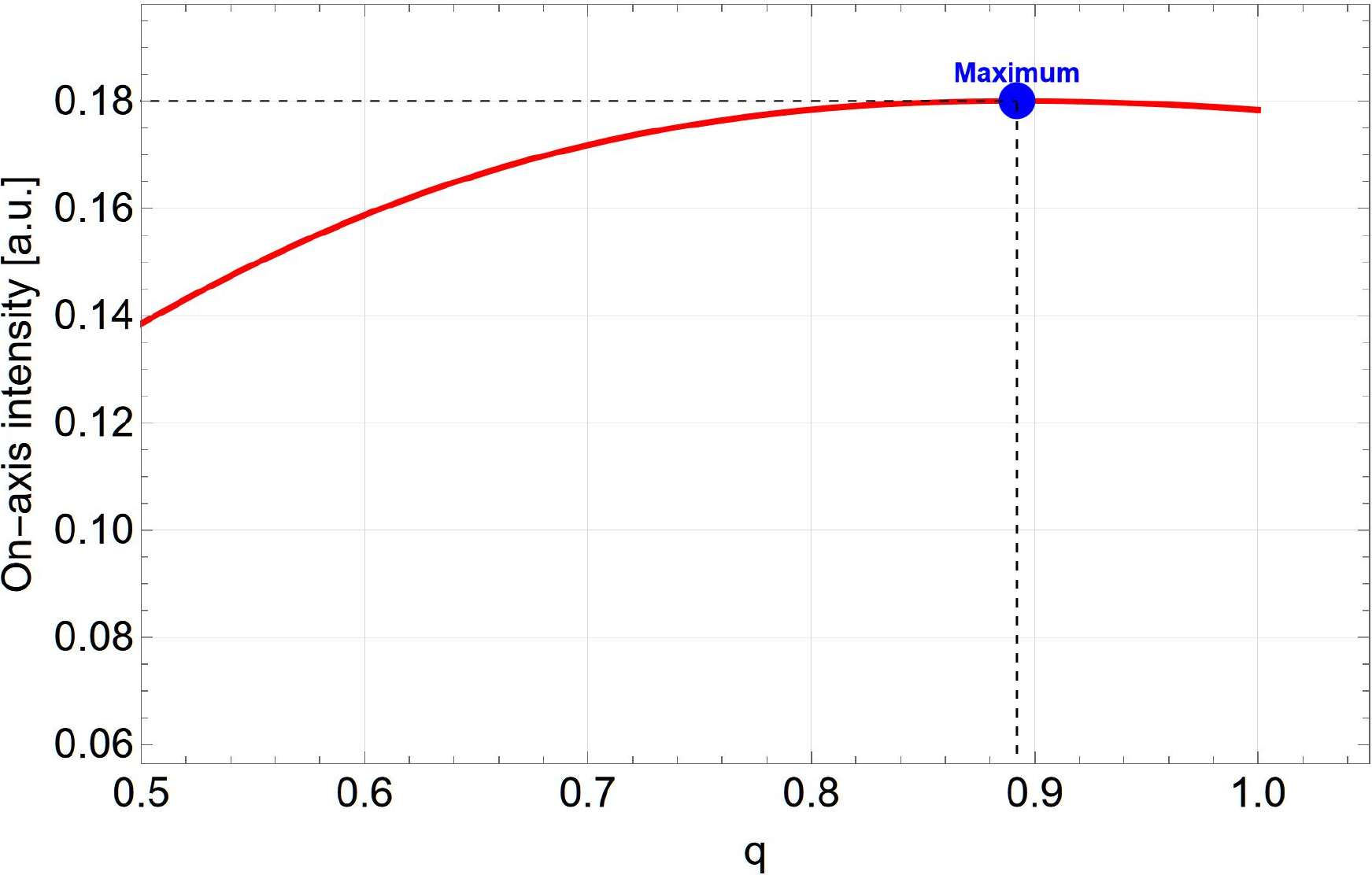}%
	\caption{Variation of the on-axis received power at the remote end with $q$. The maximum occurs at $q=0.892135$, indicating that the received beam power is maximized when the ratio of $w_0$ to $r_a$ is approximately 0.9.}
	\label{fig:tradeoff_centerPower}
  \end{figure}

Neglecting the transmitted WFE $\varOmega_a$ and the spot shift $s_r$, the far-field center can be obtained from \eqref{FraunhoferAberration1} by using the expanded base term in \eqref{U0_expanded}. At $v=0$ and a specific‌ arm length L, this gives
\begin{equation}\label{FraunhoferAberration1_center}
\begin{aligned}
E(0,0,L;q)
&=\sqrt{\frac{2P_0}{\pi w_0^2}}\frac{e^{ikz}}{i\lambda L}r_a^2
q e^{-\frac{1}{2q^2}}\pi^{\frac{3}{2}}I_{\frac{1}{2}}(\frac{1}{2q^2}),
\end{aligned}
\end{equation}
and the corresponding intensity is
\begin{equation}\label{FraunhoferAberration1_centerPower}
	I(0,0,L;q)=|E(0,0,z)|^2=\frac{2{\pi}^2P_0r_a^2}{{\lambda}^2 L^2}
	e^{-\frac{1}{q^2}}I_{\frac{1}{2}}^2(\frac{1}{2q^2}),
	\end{equation}
Here, the received power is evaluated using the far-field on-axis intensity. After omitting the constant factor in \eqref{FraunhoferAberration1_centerPower}, Fig.~\ref{fig:tradeoff_centerPower} shows the dependence of the far-field on-axis intensity on $q$.

\subsection{Influence on WFE} \label{sbse:3.2}

\begin{figure}[htbp]
	\centering
	  \includegraphics[width=0.6\textwidth]{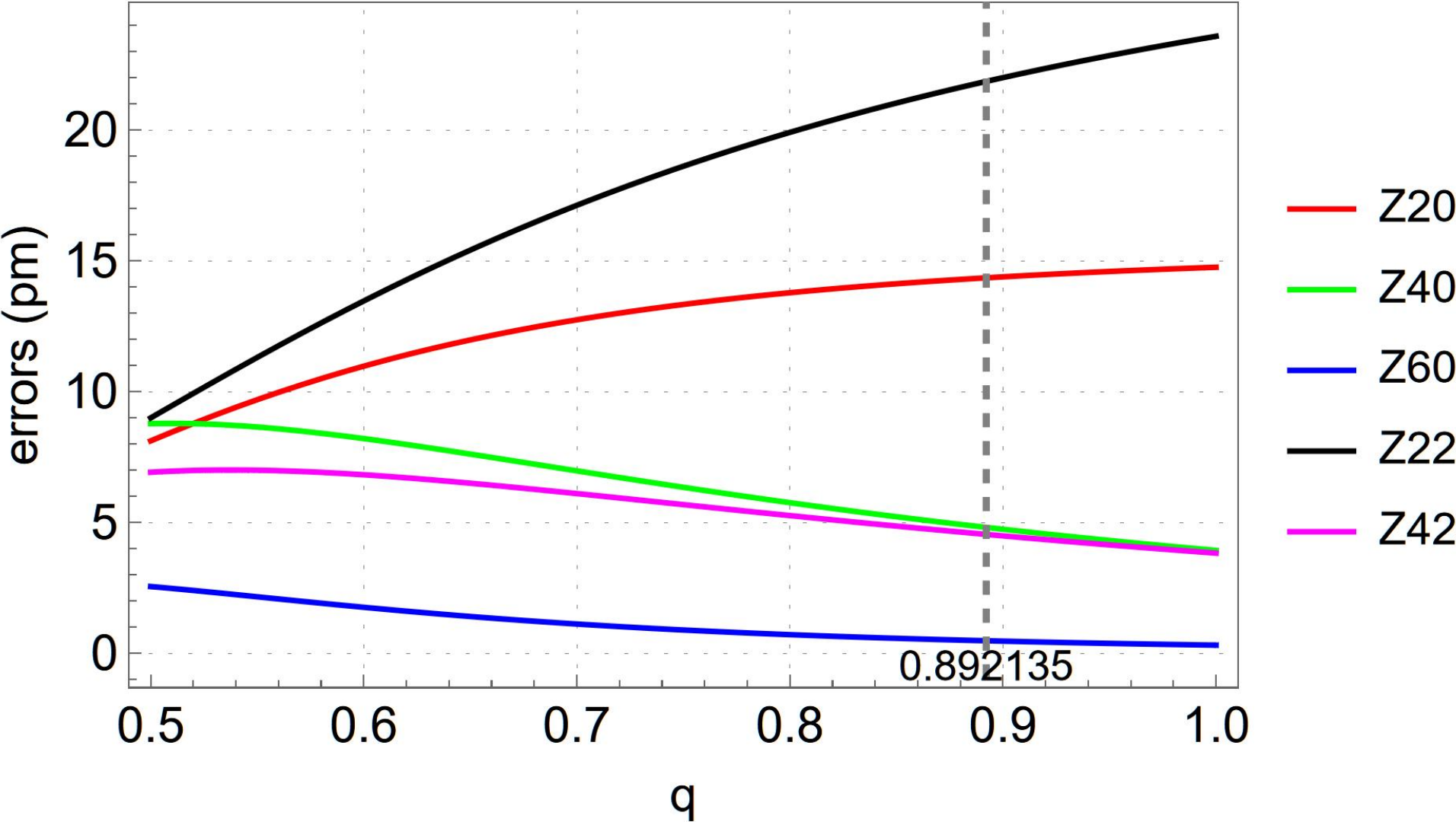}%
	\caption{Dependence of the first-order far-field WFE contributions of selected even-$n$ Zernike terms on $q$.}
	\label{fig:1order_trade}
\end{figure}

The variation of $q$ also affects the WFE. For the first-order contribution of aberrations, we calculated the even-$n$ Zernike terms among the first 21 modes that have a significant influence on the far-field WFE. The odd-$n$ terms affect the imaginary component through coupling and are therefore not considered here. The results are shown in Fig.~\ref{fig:1order_trade}. The calculation is based on the Taiji arm length $L=3\times10^9~\mathrm{m}$ and telescope aperture radius $r_a=200~\mathrm{mm}$; the following calculations also use these Taiji parameters. As $q$ increases, the contributions of $Z_{2}^0$ and $Z_{2}^{\pm2}$ increase, whereas those of the other aberration terms decrease. This trend is consistent with that reported in \cite{vinet2019lisa}. These results indicate that, in principle, the selection of $q$ also involves a trade-off with respect to the WFE.

\begin{figure}
	\centering	   
	\begin{subfigure}[b]{0.45\textwidth}
		\centering
		\caption{$Z_1^1Z^{2\beta}_{{\gamma}{'}}$}
		\includegraphics[width=1\textwidth]{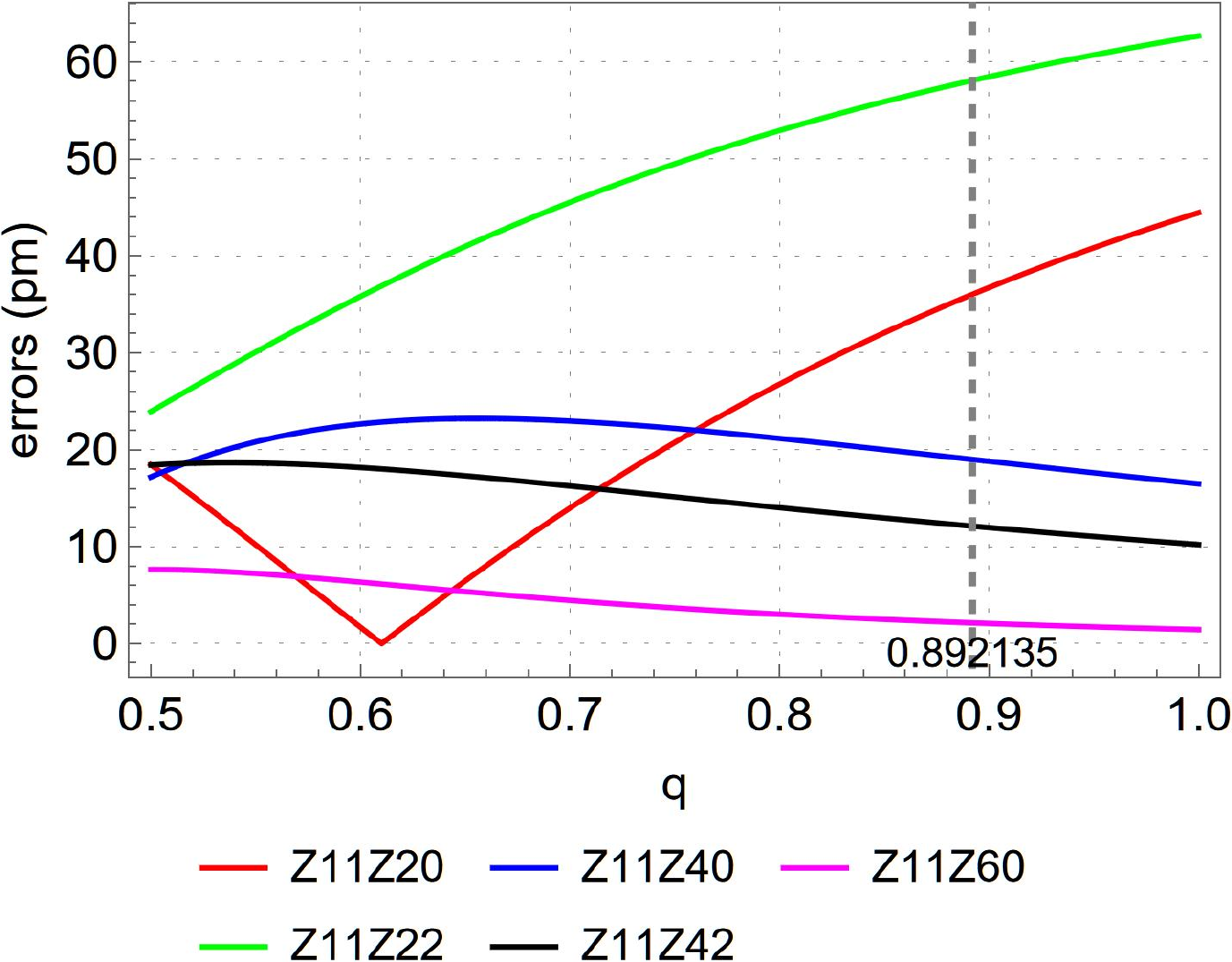}
		\label{fig:Z11_trade}
	\end{subfigure}
	\begin{subfigure}[b]{0.45\textwidth}
		\centering
		\caption{$Z_3^1Z^{2\beta}_{{\gamma}{'}}$}
		\includegraphics[width=1\textwidth]{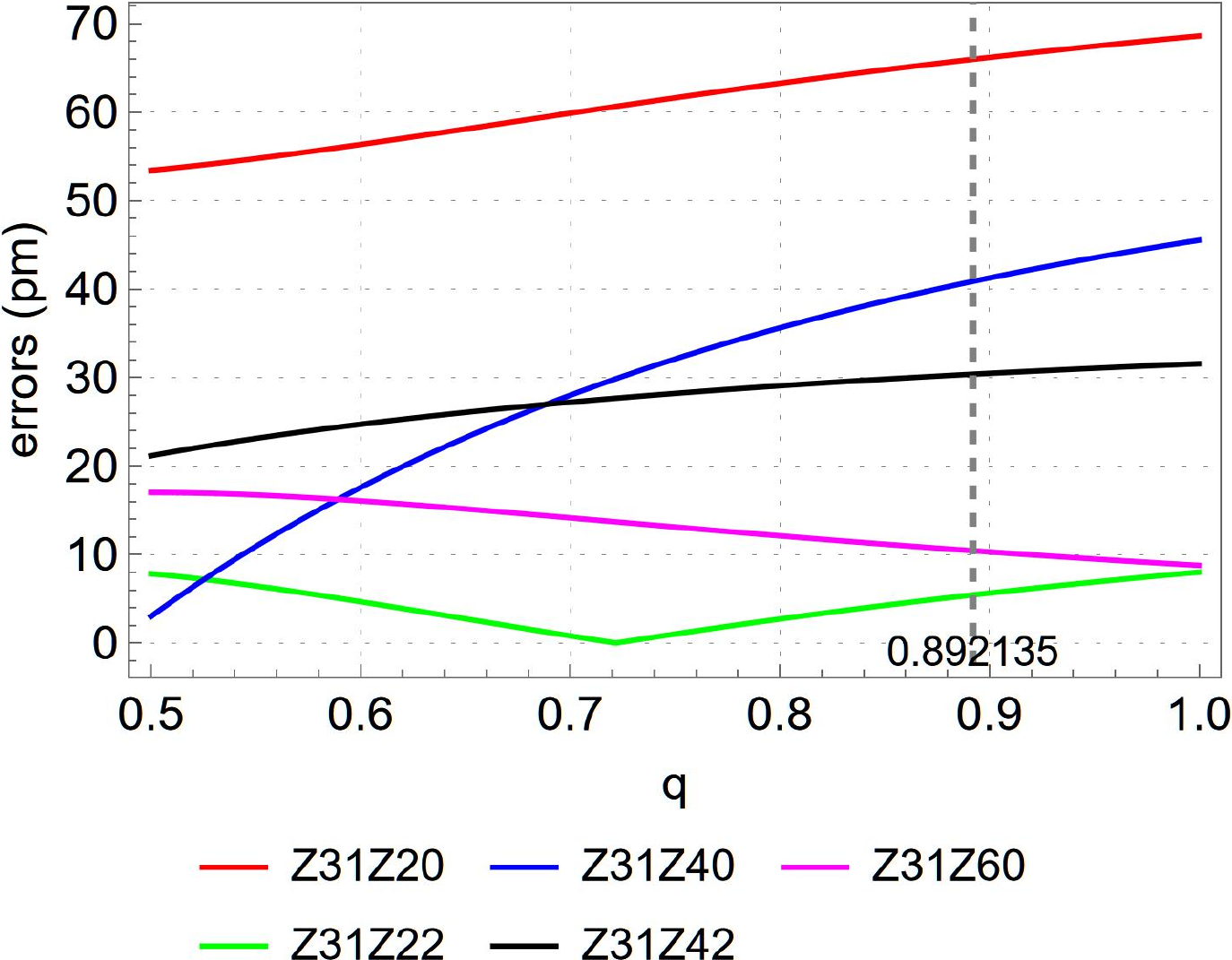}
		\label{fig:Z31_trade}
	\end{subfigure}    
		 \begin{subfigure}[b]{0.45\textwidth}
		\centering
		\caption{$Z_5^1Z^{2\beta}_{{\gamma}{'}}$}
		\includegraphics[width=1\textwidth]{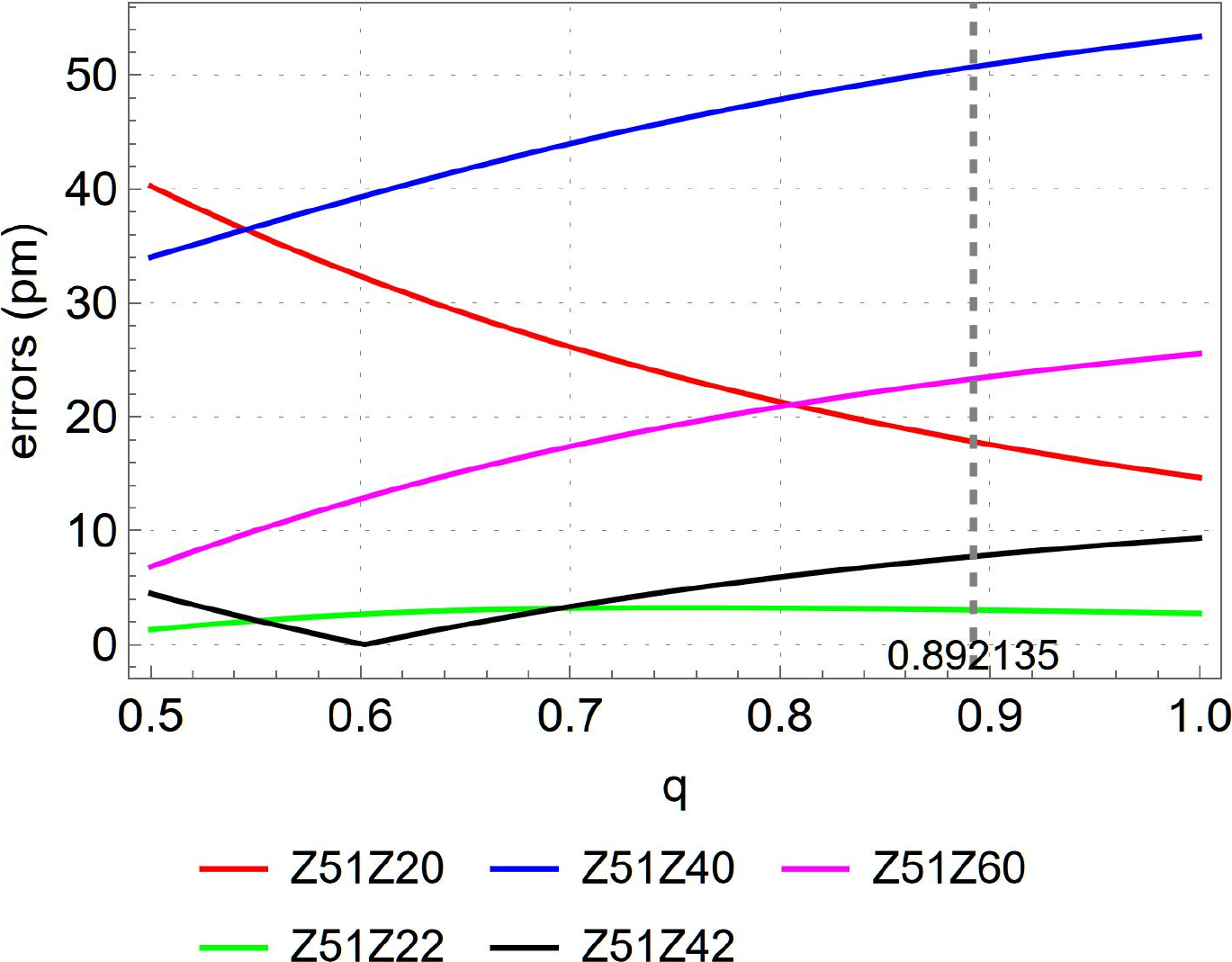}
		\label{fig:Z51_trade}        
	\end{subfigure}    
		 \begin{subfigure}[b]{0.45\textwidth}
		\centering
		\caption{Other couplings}
		\includegraphics[width=1\textwidth]{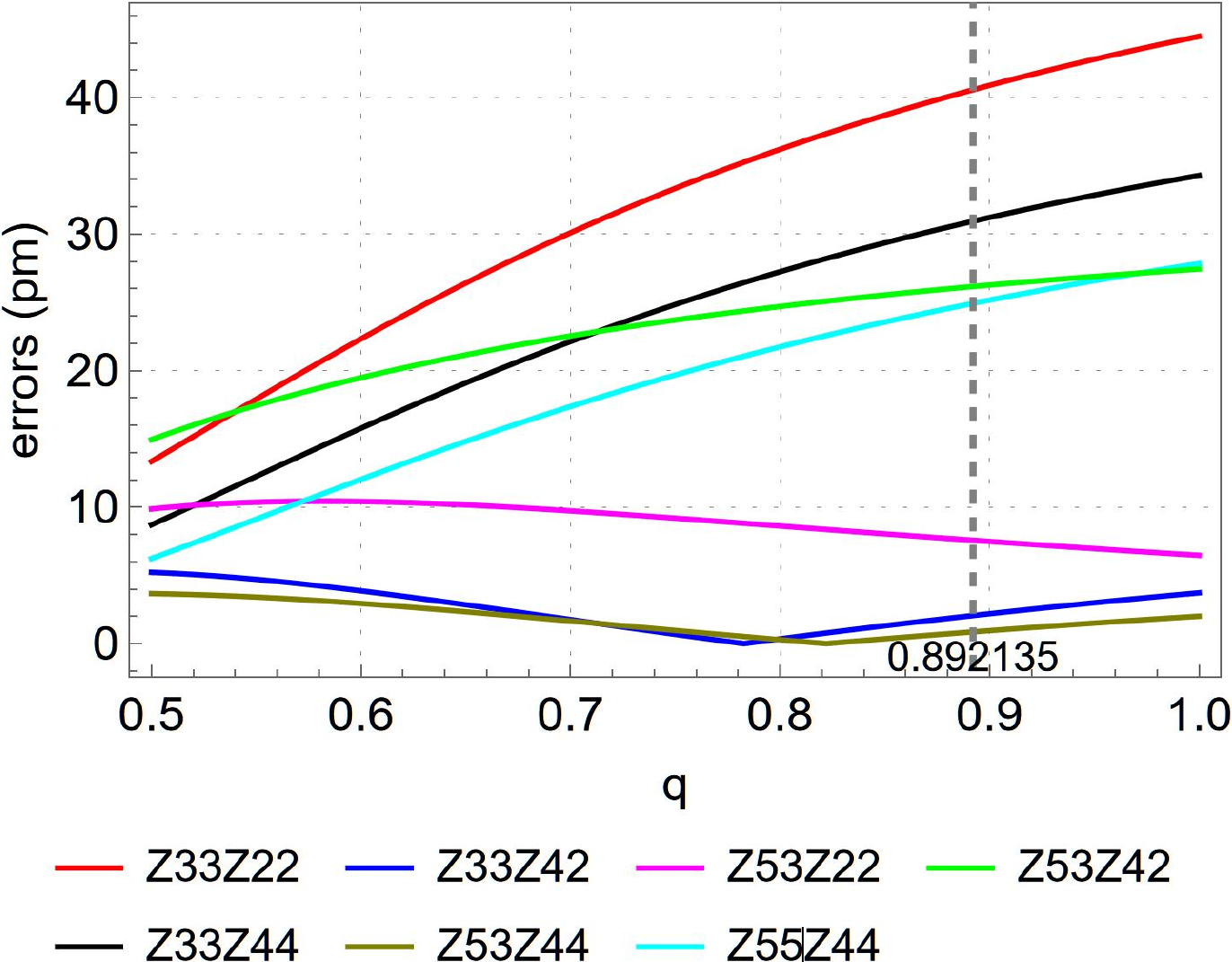}
		\label{fig:other_trade}        
	\end{subfigure}
	 \caption{Dependence of representative second-order coupled aberration terms on $q$. The subfigures show the coupling of $Z_1^1$, $Z_3^1$, $Z_5^1$, and other aberration terms with ${Z^{2\beta}_{{\gamma}{'}}}$, respectively.}
	 \label{fig:2order_trade}
	\end{figure}

Furthermore, we calculated the dependence of the second-order coupled aberrations on $q$, as shown in Fig.~\ref{fig:2order_trade}. The relationship between each coupled aberration term and $q$ exhibits complex behavior, making it difficult to establish a simple selection strategy. For example, for the odd-$n$ aberration $Z_{1}^{\pm1}$, as shown in Fig.~\ref{fig:Z11_trade}, its coupling with other even-$n$ aberrations does not show a consistent trend as $q$ varies. Similarly, for the even-$n$ aberration $Z_{2}^0$, the results in Figs.~\ref{fig:Z11_trade}, \ref{fig:Z31_trade}, and \ref{fig:Z51_trade} show that the corresponding coupling terms also do not vary monotonically with $q$.

\begin{figure}
		\centering	   
	\begin{subfigure}[b]{0.45\textwidth}
		\centering
		\caption{$q=0.8$ vs. $q=0.9$}
		\includegraphics[width=1\textwidth]{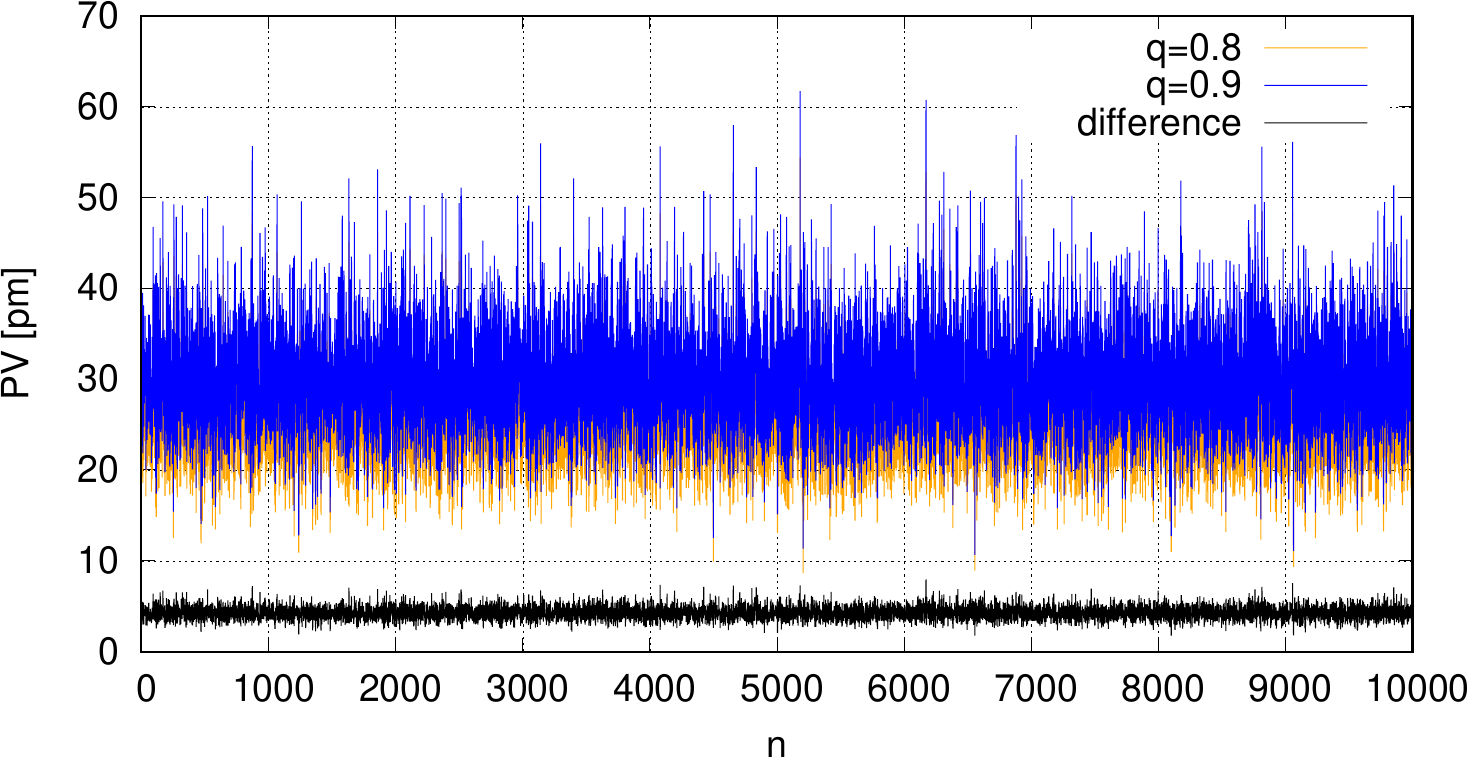}
		\label{fig:Mc_0.8-0.9}
	\end{subfigure}
	\begin{subfigure}[b]{0.45\textwidth}
		\centering
		\caption{$q=0.9$ vs. $q=1$}
		\includegraphics[width=1\textwidth]{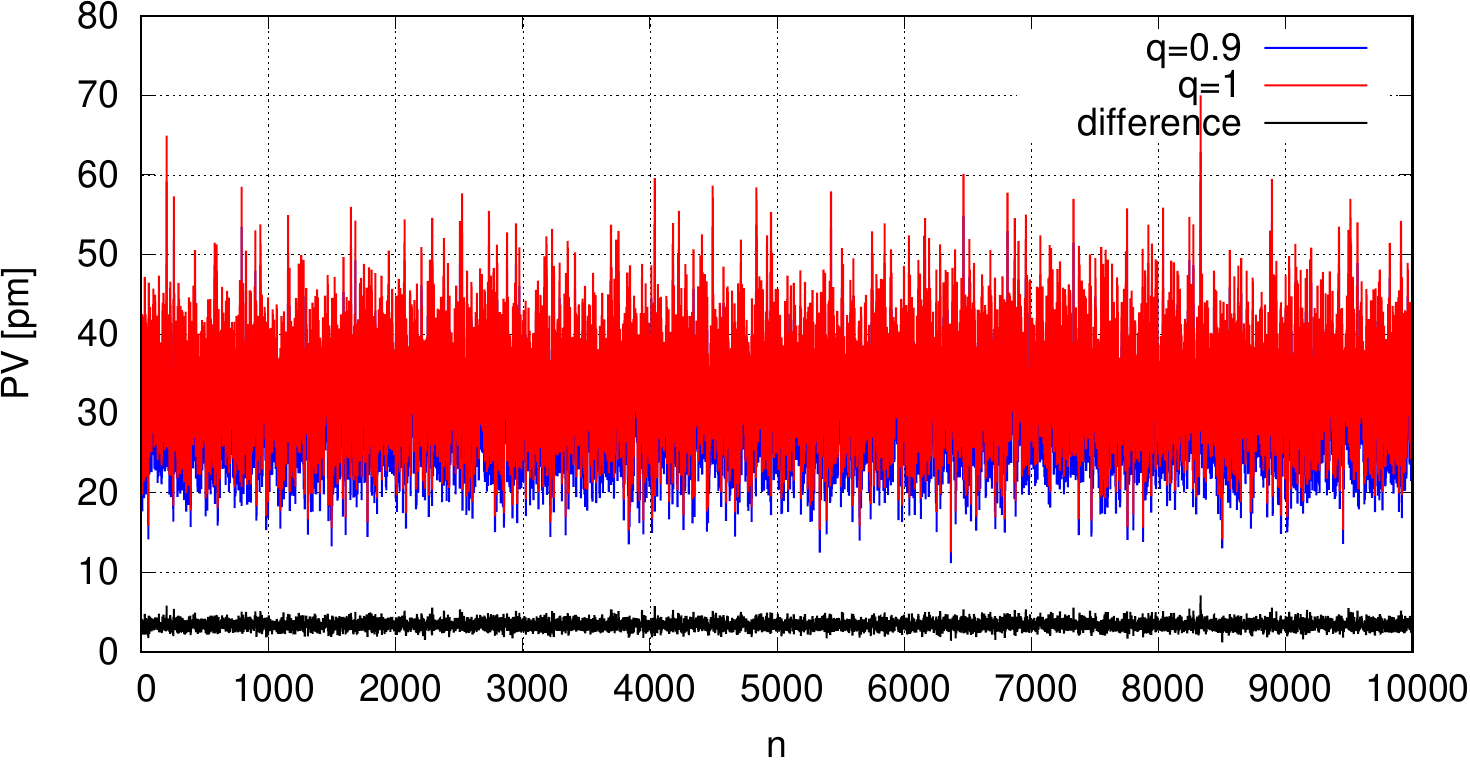}
		\label{fig:Mc_0.9-1}   
	\end{subfigure}
	 \caption{Monte Carlo comparison of the far-field WFE for different values of $q$. The left panel compares $q=0.8$ with $q=0.9$, and the right panel compares $q=0.9$ with $q=1$.}
		 \label{fig:Mc}
\end{figure}

Given this complexity, we adopt a Monte Carlo method. In each trial, the coefficients of the first 21 Zernike aberration terms are independently sampled with equal weight from the same uniform interval. These coefficients are then combined to form an initial WFE map, which is subsequently normalized so that its P-V value is fixed at $\lambda/20$. The far-field WFE levels are then compared for different values of $q$, as shown in Fig.~\ref{fig:Mc}. 

The results show that, among the $10^4$ Monte Carlo trials, no case was found in which the far-field WFE for $q=0.9$ exceeded that for $q=1$. Similarly, no case was found in which the far-field WFE for $q=0.8$ exceeded that for $q=0.9$. The mean far-field WFE was then calculated over the $10^4$ trials in each comparison. In the left panel of Fig.~\ref{fig:Mc}, the mean values for $q=0.8$ and $q=0.9$ are 25.5782 pm and 29.8319 pm, respectively. The former is, on average, approximately 14$\%$ lower than the latter. In the right panel, the corresponding values for $q=0.9$ and $q=1$ are 29.7208 pm and 33.0843 pm, respectively. The former is, on average, approximately 10$\%$ lower than the latter. Therefore, for most initial WFE conditions with $0.8\leq{q}\leq1$, reducing the beam-waist-to-aperture ratio can effectively decrease the resulting far-field WFE. This estimate is based on a fully random distribution of the first 21 aberration terms. A more refined evaluation should further consider the relative weights of different aberration terms generated by a specific telescope.

\section{Normalized lateral spot shift ratio \textit{s\textsubscript{r}}}\label{se:4}

\subsection{Spot shift term $U_2(r, \psi, z)$} \label{sbse:4.1}
We use $U_0(r, \psi, z)$ and $U_2(r, \psi, z)$ to investigate the effect of the spot-shift term on the far-field wavefront error. Assuming that no transmitted WFE is present, i.e., $\varOmega_a(\rho,\theta)=0$, we obtain:

\begin{equation}\label{OnlyShift}
U(r, \psi, z)=qe^{-\frac{1}{2q^2}}{\pi}^{\frac{1}{2}}(2\pi)\left[\left\{{\sigma}_0\frac{J_1(v)}{v}+{\tau}_0\frac{J_3(v)}{v}\right\}-i2s_r/q\cos(\psi-{\theta}_0)\left\{{\eta}_0^1\frac{J_2(v)}{v}+{\zeta}_0^1\frac{J_4(v)}{v}\right\}\right]
\end{equation}
Therefore, when the beam-spot center is laterally shifted, an additional far-field wavefront error is introduced even in the absence of coupling with the transmitted WFE. Regarding the amplitude, the spot-shift term contributes only to the imaginary part of the expression. Owing to the large difference in magnitude between the imaginary and real parts, this term has a negligible effect on the beam amplitude. The amplitude and the additional WFE are given by:
\begin{subequations}\label{OnlyShift_2}
\begin{align}
 \label{OnlyShift_Amp}
&|E(r, \psi, z)|\approx Re\{E(r, \psi, z)\}=\sqrt{\frac{2P_0}{\pi}}\frac{{r_a}}{\lambda zq}{e^{-s_r^2}}Re\left\{U(r, \psi, z)\right\},\\
 \label{OnlyShift_WFE}
&WFE_s(v, \psi;s_r,{\theta}_0)=-2\frac{s_r}{q}\cos(\psi-{\theta}_0)\frac{{\eta}_0^1J_2(v)+{\zeta}_0^1J_4(v)}{{\sigma}_0J_1(v)+{\tau}_0J_3(v)}.
\end{align}
\end{subequations}
where
\begin{gather*}\label{OnlyShift_WFEco}
	\left({\eta}_0^1,\;{\zeta}_0^1\right)=\left(I_{\frac{1}{2}}-I_{\frac{3}{2}},\;2(I_{\frac{3}{2}}-I_{\frac{5}{2}})\right).
\end{gather*}

\begin{figure}[htbp]
	\centering
	  \includegraphics[width=0.5\textwidth]{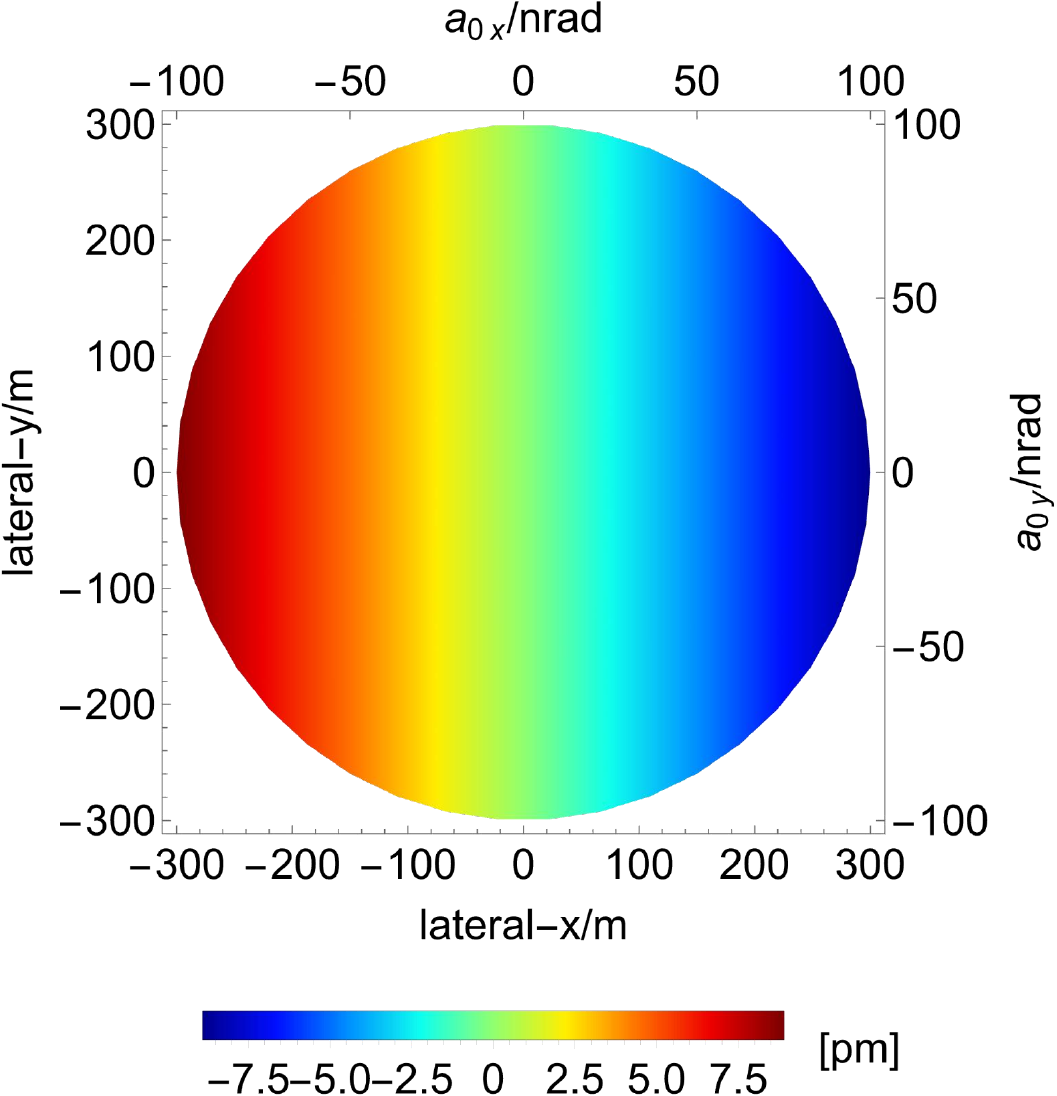}%
	\caption{Far-field WFE induced by the spot shift for $s_r=0.001$.}
	\label{fig:sr0.001WFE}
\end{figure}

We take the value of $q$ that maximizes the received optical power, namely $q=0.892135$. The expression of far-field WFE can then be approximated as follows:
\begin{equation}\label{OnlyShift_WFE_approx}
WFE_s(\alpha, \psi;s_r,{\theta}_0)=-89.1528s_r \cos(\psi-{\theta}_0)\cdot\alpha
\end{equation}
where $\alpha$ is the jitter angle. It can be observed that the overall level of the far-field wavefront error increases with the shift ratio $s_r$ and decreases with the ratio of the beam waist to the aperture radius $q$. We calculated the far-field WFE for $s_r=0.001$, as shown in Fig.~\ref{fig:sr0.001WFE}. The phase-angle coupling coefficient is approximately $0.0892~\mathrm{pm/nrad}$. This value is already close to the requirement of $0.1~\mathrm{pm/nrad}$ for the total phase-angle coupling coefficient of the far-field TTL coupling noise.

For $s_r=0.001$, taking the Taiji telescope as an example, the aperture radius is $200~\mathrm{mm}$ and the magnification is $100\times$. This corresponds to a lateral displacement of only $2~\mu\mathrm{m}$ for the laser beam at the entrance pupil. Such a small displacement imposes stringent requirements on the optical assembly.

\subsection{Coupling between the spot shift and the transmitted wavefront error term $U_3(r, \psi, z)$} \label{sbse:4.2}
We next consider the fourth term in \eqref{U_Expand}. If only the first-order coupling between the spot-center offset and the transmitted wavefront error is retained, the resulting integral takes a form similar to that for the coupling between tilt aberration and the other aberration terms. Since different aberration-coupling terms contribute differently to the far-field wavefront error, only terms with non-negligible contributions need to be retained\cite{tao2025approximate}. Therefore, among the first 21 Zernike aberration terms, for even-azimuthal-order aberrations, only the couplings between the spot-center offset and $Z_2^0$, $Z_2^{\pm2}$, $Z_4^0$, $Z_4^{\pm2}$, and $Z_6^0$ need to be considered. For odd-azimuthal-order aberrations, the couplings with $Z_1^{\pm1}$, $Z_3^{\pm1}$, and $Z_5^{\pm1}$ should be taken into account.

Similarly, we derived the expressions for the corresponding coupling terms, where $S^{-1}$ denotes the Y-direction lateral shift, $S^{1}$ denotes the X-direction lateral shift, and $S^{a}Z_n^m$ denotes the associated coupling term:
\begin{equation}\label{SZn0}
	\begin{pmatrix}
			S^{1}Z_{n{'}}^{0} \\
			S^{-1}Z_{n{'}}^{0} \\
		\end{pmatrix}
		(r, \psi, z)=2\frac{s_r}{q}e^{-\frac{1}{2q^2}}{\pi}^{\frac{1}{2}}(2\pi)\left\{
	\begin{pmatrix}
			\cos\psi \\
			\sin\psi \\
		\end{pmatrix}
	\left[{\eta}_{;n}^{\pm1;0}\frac{J_2(v)}{v}+{\zeta}_{;n}^{\pm1;0}\frac{J_4(v)}{v}\right]\right\},
\end{equation}
\begin{equation}\label{SZnpm2}
\begin{aligned}
	\begin{pmatrix}
			S^{1}Z_{n{'}}^{2} &  S^{1}Z_{n{'}}^{-2}\\
			S^{-1}Z_{n{'}}^{2} & S^{-1}Z_{n{'}}^{-2}\\
		\end{pmatrix}
		(r, \psi, z)=2\frac{s_r}{q}e^{-\frac{1}{2q^2}}{\pi}^{\frac{1}{2}}(2\pi)\left\{\frac{1}{2}
	\begin{pmatrix}
			\cos\psi &  \sin\psi\\
			-\sin\psi & \cos\psi\\
		\end{pmatrix}
			\left[{{\eta}}_{;n}^{\pm1;\pm2}\frac{J_2(v)}{v}+{{\zeta}}_{;n}^{\pm1;\pm2}\frac{J_4(v)}{v}\right]\right\},	
\end{aligned}
\end{equation}
\begin{equation}\label{SZn1}
\begin{aligned}
	&\begin{pmatrix}
			S^{1}Z_{n{'}}^{1} \\
			S^{-1}Z_{n{'}}^{-1} \\
		\end{pmatrix}
		(r, \psi, z)=2\frac{s_r}{q}e^{-\frac{1}{2q^2}}{\pi}^{\frac{1}{2}}(2\pi)\\
		&\left\{-\frac{i}{2}\left[{{\eta}_1}_{;n}^{\pm1;\pm1}\frac{J_1(v)}{v}+{{\zeta}_1}_{;n}^{\pm1;\pm1}\frac{J_3(v)}{v}\right]+\frac{i}{2}
	\begin{pmatrix}
			\cos2\psi \\
			-\cos2\psi \\
		\end{pmatrix}
	\left[{{\eta}_2}_{;n}^{\pm1;\pm1}\frac{J_1(v)}{v}+{{\zeta}_2}_{;n}^{\pm1;\pm1}\frac{J_3(v)}{v}\right]\right\},
\end{aligned}
\end{equation}
The coupling coefficients of these terms are listed in Table \ref{S1CouplingCoefficients}.
\begin{table*}[htbp]
	\abovetopsep=0pt
	\aboverulesep=0pt
	\belowrulesep=0pt
	\belowbottomsep=0pt
	\centering
	\small
	\setlength{\tabcolsep}{3mm}
	\renewcommand\arraystretch{1.9}
	\begin{tabular}{m{1cm}|m{2.3cm}|m{8.8cm}}
		\toprule[1.5pt]
		Term & Coefficients & Values \\
		\midrule[1.2pt]
		\multicolumn{3}{c}{Even-order aberration couplings} \\
		\midrule[0.9pt]
		$S^{\pm1}Z_2^{0}$ & $({{\eta}}_{;2}^{\pm1;0},\;{{\zeta}}_{;2}^{\pm1;0})$ &
		\makecell[l]{$a_2^0(0.333333L_0+0.333333L_1+0.133333L_2,$\\$-0.666667L_0-0.266667L_1-0.266667L_2-0.171429L_3)$} \\
		\hline
		$S^{\pm1}Z_4^{0}$ & $({{\eta}}_{;4}^{\pm1;0},\;{{\zeta}}_{;4}^{\pm1;0})$ &
		\makecell[l]{$a_4^0(0.133333L_1+0.2L_2+0.0857143L_3,$\\$-0.4L_0-0.266667L_1-0.114286L_2-0.171429L_3)$} \\
		\hline
		$S^{\pm1}Z_2^{\pm2}$ & $({{\eta}}_{;2}^{\pm1;\pm2},\;{{\zeta}}_{;2}^{\pm1;\pm2})$ &
		\makecell[l]{$a_2^{\pm2}(0.666667L_0+0.333333L_1+0.0666667L_2,$\\$-0.333333L_0-0.466667L_1-0.333333L_2-0.0857143L_3)$} \\
		\hline
		$S^{\pm1}Z_4^{\pm2}$ & $({{\eta}}_{;4}^{\pm1;\pm2},\;{{\zeta}}_{;4}^{\pm1;\pm2})$ &
		\makecell[l]{$a_4^{\pm2}(0.2L_1+0.2L_2+0.0571429L_3,$\\$-0.6L_0-0.2L_1-0.142857L_2-0.2L_3)$} \\
		\midrule[1.2pt]
		\multicolumn{3}{c}{Odd-order aberration couplings} \\
		\midrule[0.9pt]
		$S^{\pm1}Z_1^{\pm1}$ &
		\makecell[c]{$({{\eta}}_{1;1}^{\pm1;\pm1},\;{{\zeta}}_{1;1}^{\pm1;\pm1})_1$\\$({{\eta}}_{2;1}^{\pm1;\pm1},\;{{\zeta}}_{2;1}^{\pm1;\pm1})_2$} &
		\makecell[l]{$a_1^{\pm1}(0.5L_0+0.166667L_1,\;-0.5L_0-0.5L_1-0.2L_2)_1$\\$a_1^{\pm1}(0,\;L_0+0.5L_1+0.1L_2)_2$} \\
		\hline
		$S^{\pm1}Z_3^{\pm1}$ &
		\makecell[c]{$({{\eta}}_{1;3}^{\pm1;\pm1},\;{{\zeta}}_{1;3}^{\pm1;\pm1})_1$\\$({{\eta}}_{2;3}^{\pm1;\pm1},\;{{\zeta}}_{2;3}^{\pm1;\pm1})_2$} &
		\makecell[l]{$a_3^{\pm1}(0.166667L_1+0.1L_2,\;-0.5L_0-0.2L_1-0.2L_2-0.128571L_3)_1$\\$a_3^{\pm1}(0,\;0.25L_0+0.35L_1+0.25L_2+0.0642857L_3)_2$} \\
		\hline
		$S^{\pm1}Z_5^{\pm1}$ &
		\makecell[c]{$({{\eta}}_{1;5}^{\pm1;\pm1},\;{{\zeta}}_{1;5}^{\pm1;\pm1})_1$\\$({{\eta}}_{2;5}^{\pm1;\pm1},\;{{\zeta}}_{2;5}^{\pm1;\pm1})_2$} &
		\makecell[l]{$a_5^{\pm1}(0.1L_2+0.0714286L_3,\;-0.2L_1-0.128571L_2-0.128571L_3)_1$\\$a_5^{\pm1}(0,\;0.1L_1+0.214286L_2+0.171429L_3)_2$} \\
		\bottomrule[1.5pt]
	\end{tabular}
	\caption{Coupling coefficients for the spot-shift term $S^{\pm1}$ and the selected aberration terms in \eqref{SZn0}, \eqref{SZnpm2}, and \eqref{SZn1}. Each entry is multiplied by the corresponding Zernike coefficient. Subscripts $_1$ and $_2$ denote the first and second angular channels in \eqref{SZn1}.}
	\label{S1CouplingCoefficients}
\end{table*}

Combining the results from the preceding subsections with those obtained here, we categorize each term according to its contribution to the real and imaginary parts of the far-field wavefront error, as shown in Table \ref{ReAndIm}.
\begin{table*}
\abovetopsep=0pt
\aboverulesep=0pt
\belowrulesep=0pt
\belowbottomsep=0pt
	\begin{center}
		\begin{tabular}{c|c|c}
			\toprule[1.5pt]
			  & Real part of $U(r, \psi, z)$ &  Imaginary part of $U(r, \psi, z)$\\ 
		\midrule[1.5pt]
  First order & ${Z^{2\alpha+1}_{\gamma}}$ & ${Z^{2\beta}_{{\gamma}{'}}}\quad\boldsymbol{S}^{\pm1}$\\
  Second order & ${Z^{2{\alpha}_1+1}_{{\gamma}_1}}{Z^{2{\alpha}_2+1}_{{\gamma}_2}}\quad{Z^{2{\beta}_1}_{{{\gamma}_1}{'}}}{Z^{2{\beta}_2}_{{{\gamma}_2}{'}}}\quad\boldsymbol{{S}^{\pm1}{Z^{2\beta}_{{\gamma}{'}}}}$ & ${Z^{2\alpha+1}_{\gamma}}{Z^{2\beta}_{{\gamma}{'}}}\quad\boldsymbol{{S}^{\pm1}{Z^{2\alpha+1}_{\gamma}}}$\\
			\bottomrule[1.5pt]
		\end{tabular}
		\caption{Classification of aberration terms and their higher-order coupling terms according to their contributions to the real and imaginary parts of $U(r, \psi, z)$, where ${Z^{2\alpha+1}_{\gamma}}$ denotes the Zernike terms $Z^m_n$ with odd $m$, ${Z^{2\beta}_{{\gamma}{'}}}$ denotes those with even $m$, and ${S^{\pm1}}$ denotes the spot-shift term.}
\label{ReAndIm}
	\end{center}
\end{table*}
Although lateral displacement has an expression similar to that of tilt aberration, its contribution and the contributions of its second-order coupling terms act on the real and imaginary components in the opposite way to those of tilt aberration.

\begin{figure}
	\centering	   
	\begin{subfigure}[b]{0.3\textwidth}
		\centering
		\caption{$S^1Z_1^1$}
		\includegraphics[width=1\textwidth]{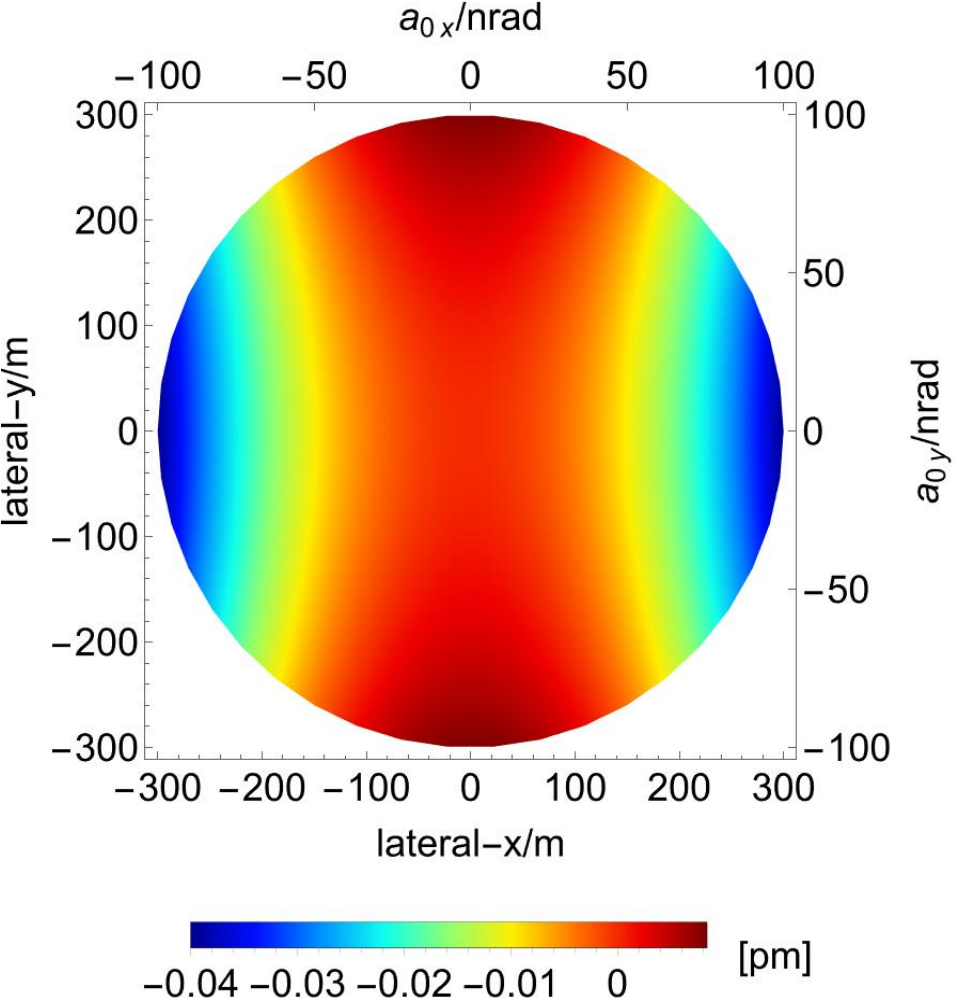}
		\label{fig:S1Z11}
	\end{subfigure}
	\begin{subfigure}[b]{0.3\textwidth}
		\centering
		\caption{$S^1Z_3^1$}
		\includegraphics[width=1\textwidth]{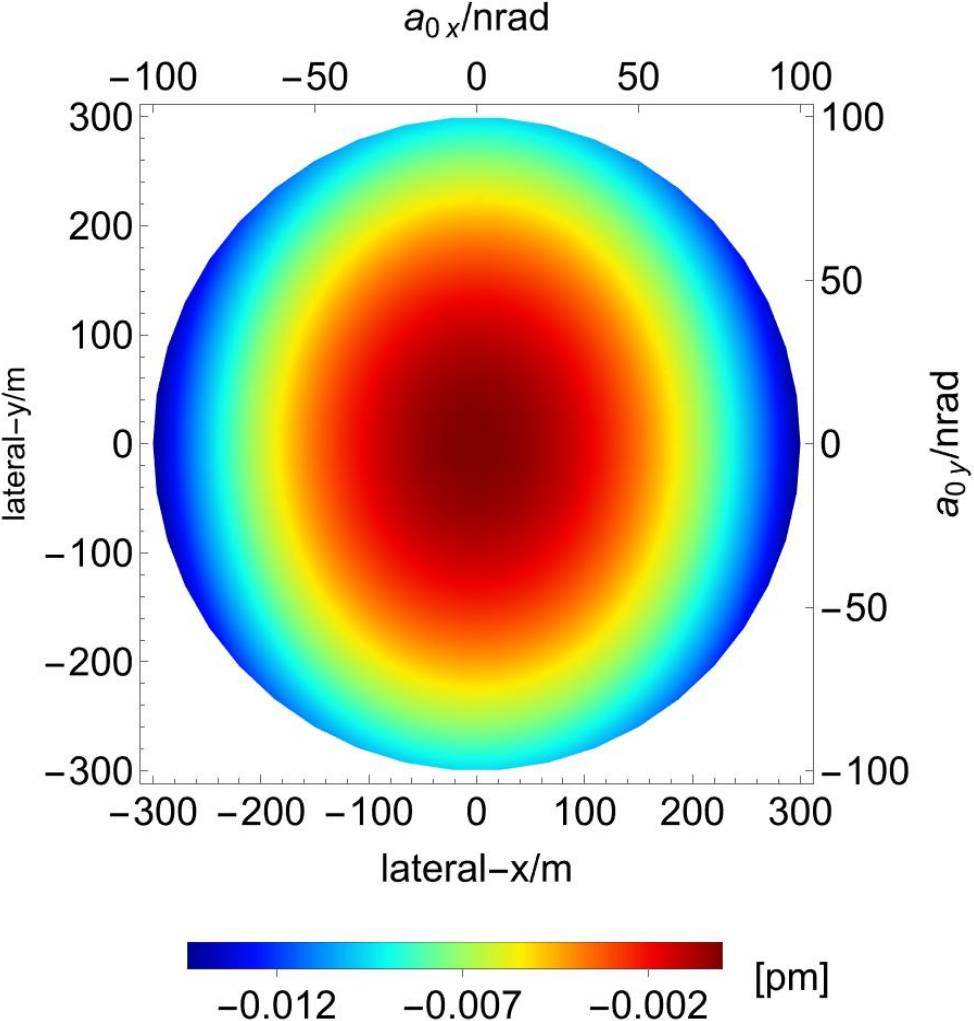}
		\label{fig:S1Z31}
	\end{subfigure}    
		 \begin{subfigure}[b]{0.3\textwidth}
		\centering
		\caption{$S^1Z_5^1$}
		\includegraphics[width=1\textwidth]{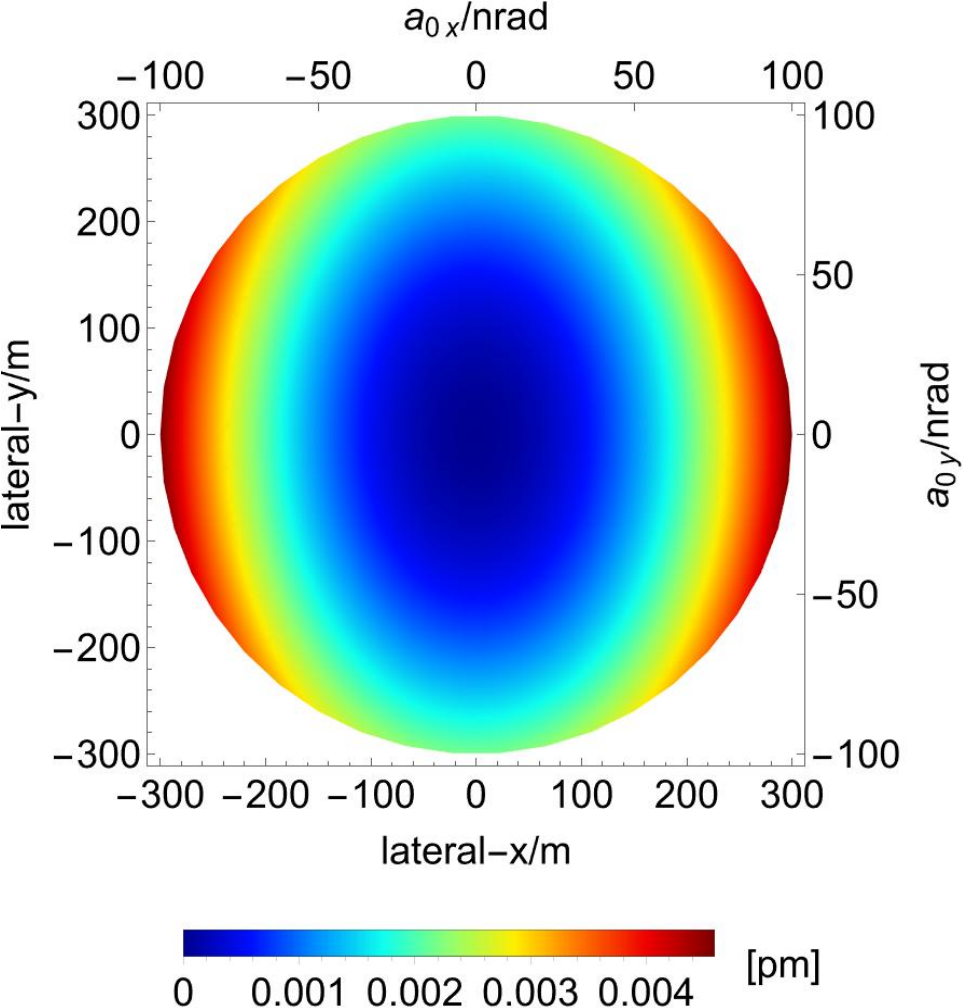}
		\label{fig:S1Z51}        
	\end{subfigure}
		\begin{subfigure}[b]{0.3\textwidth}
		\centering
		\caption{$S^1Z_3^3$}
		\includegraphics[width=1\textwidth]{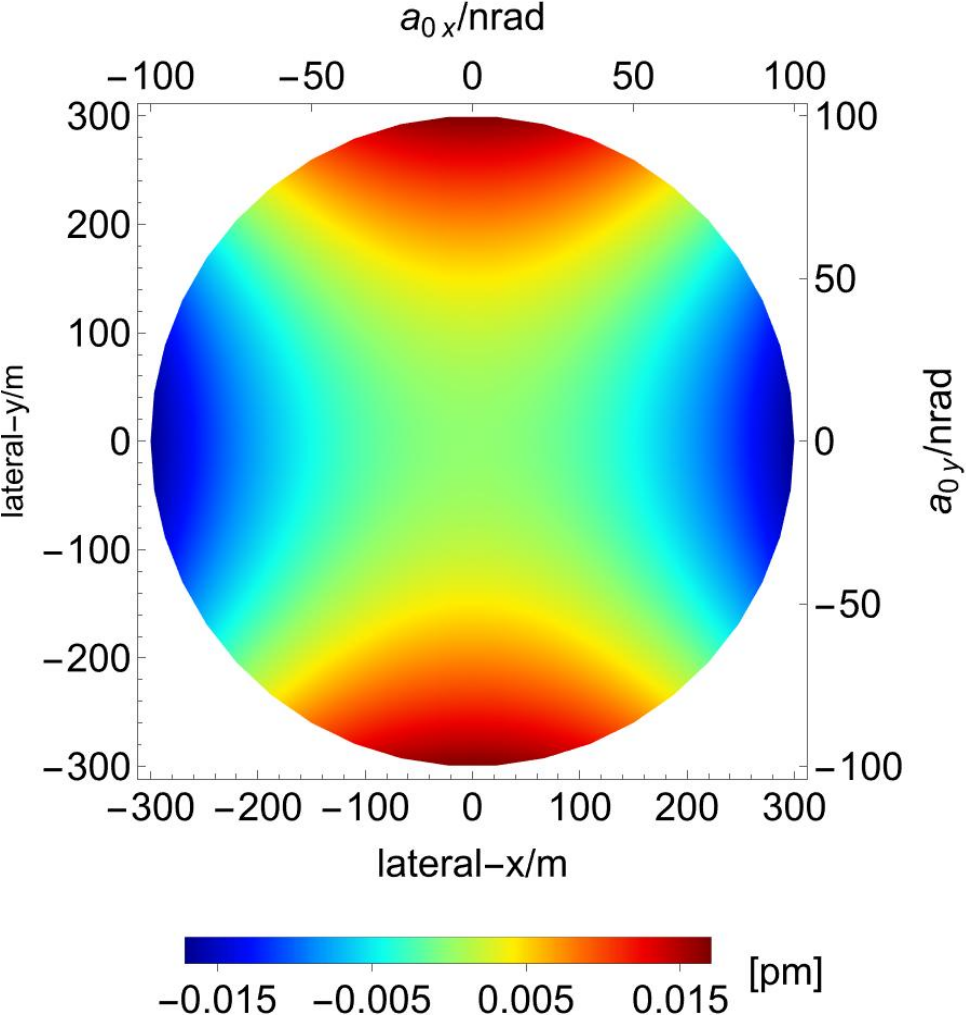}
		\label{fig:S1Z33}        
	\end{subfigure}	    
		 \begin{subfigure}[b]{0.3\textwidth}
		\centering
		\caption{$S^1Z_5^3$}
		\includegraphics[width=1\textwidth]{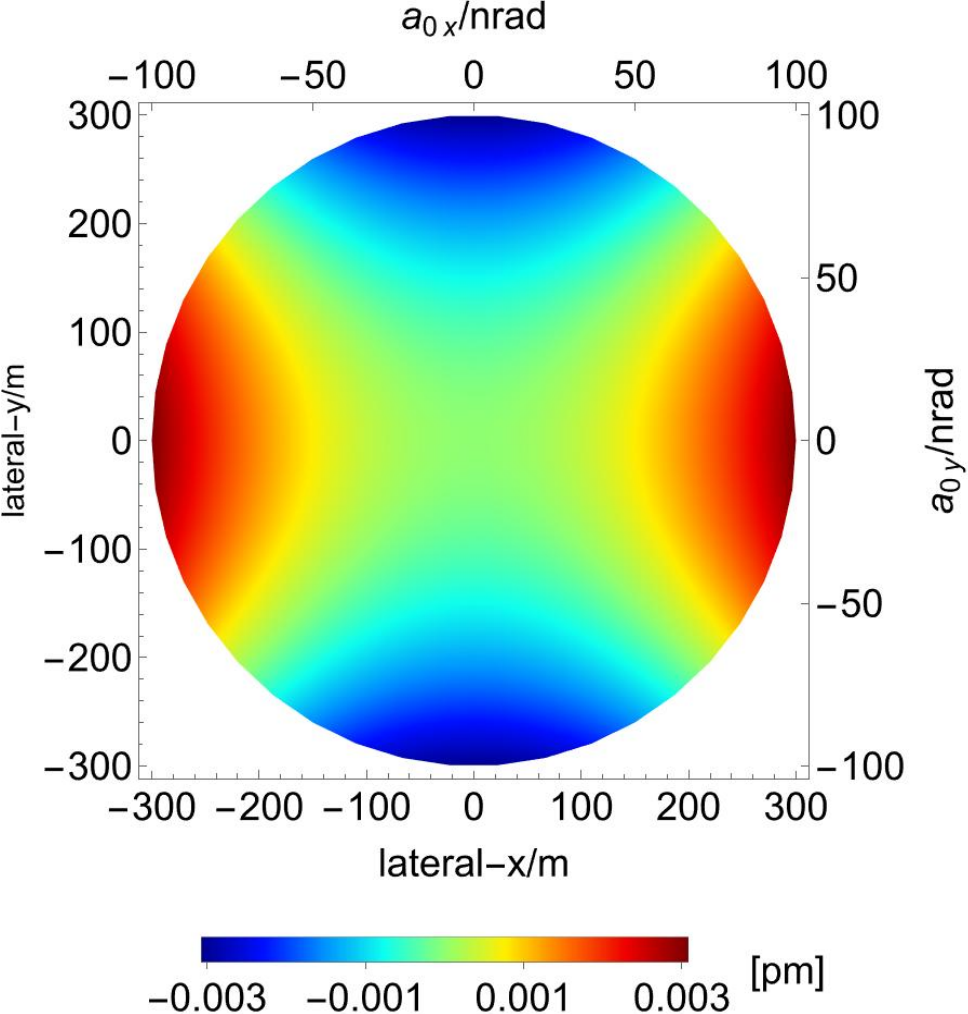}
		\label{fig:S1Z53}        
	\end{subfigure}
	 \caption{Far-field WFE induced by the second-order coupling between $S^1$ and the selected odd-$n$ aberration terms. The transmitted WFE of each aberration is constrained to $\lambda/10$, i.e., 0.314159. For all terms of the form $Z_{\gamma}^{\pm{m}}$, only $Z_{\gamma}^{m}$ is shown.}
	 \label{fig:2order_WFEodd}
\end{figure}

\begin{figure}
		\centering	   
		\begin{subfigure}[b]{0.3\textwidth}
			\centering
			\caption{$S^1Z_2^0$}
			\includegraphics[width=1\textwidth]{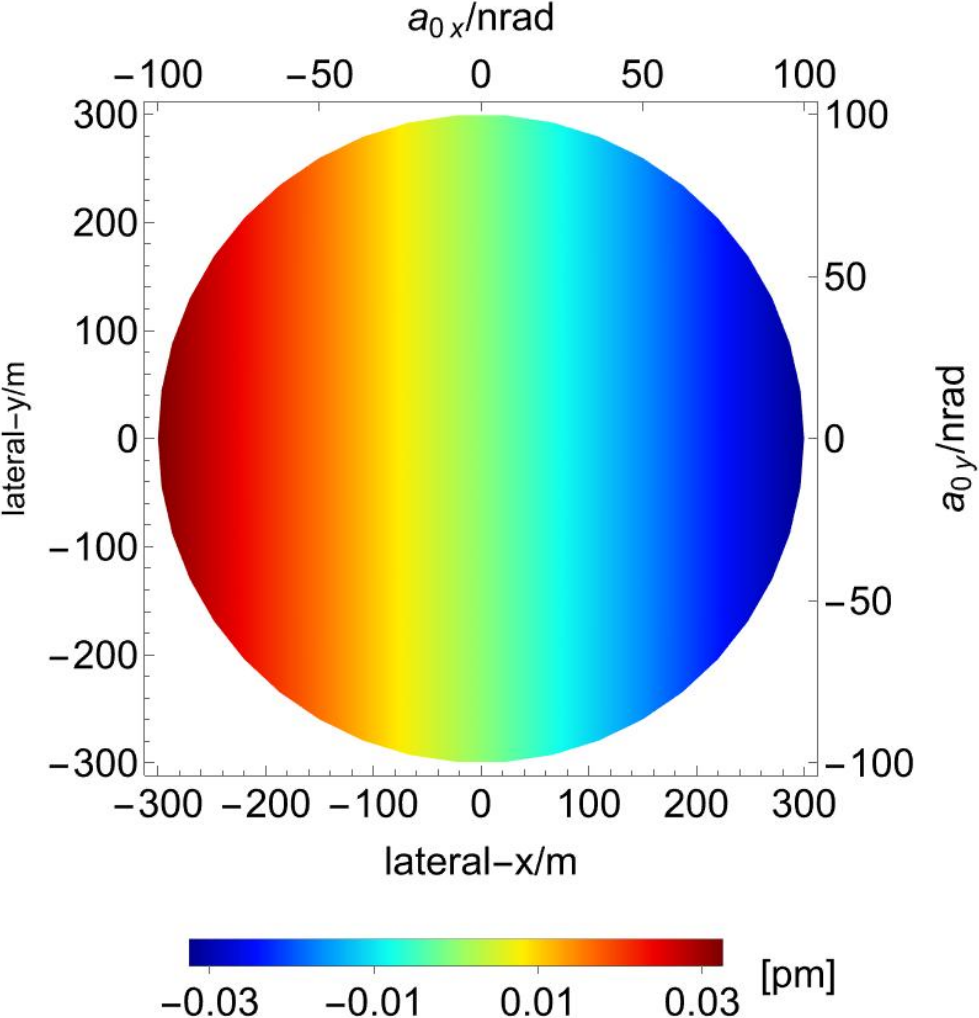}
			\label{fig:S1Z20}
		\end{subfigure}
		\begin{subfigure}[b]{0.3\textwidth}
			\centering
			\caption{$S^1Z_4^0$}
			\includegraphics[width=1\textwidth]{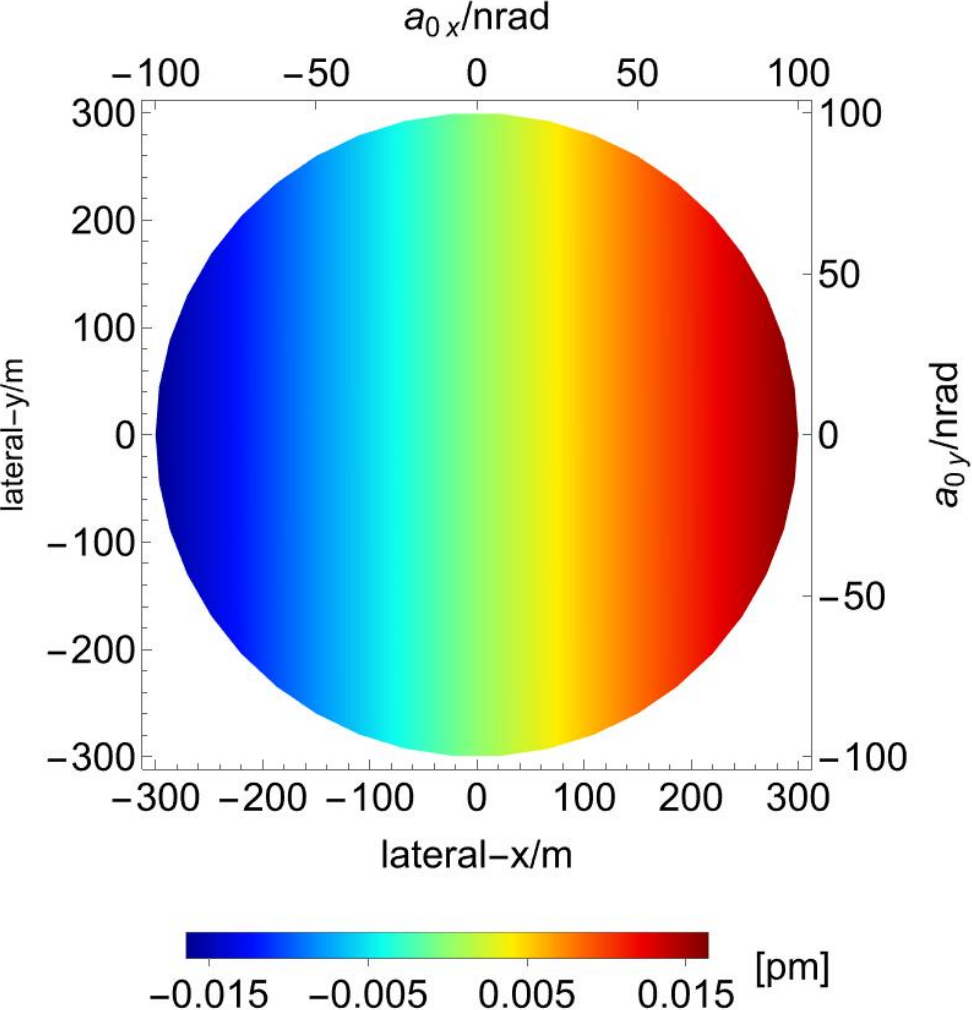}
			\label{fig:S1Z40}
		\end{subfigure}
		\begin{subfigure}[b]{0.3\textwidth}
			\centering
			\caption{$S^1Z_6^0$}
			\includegraphics[width=1\textwidth]{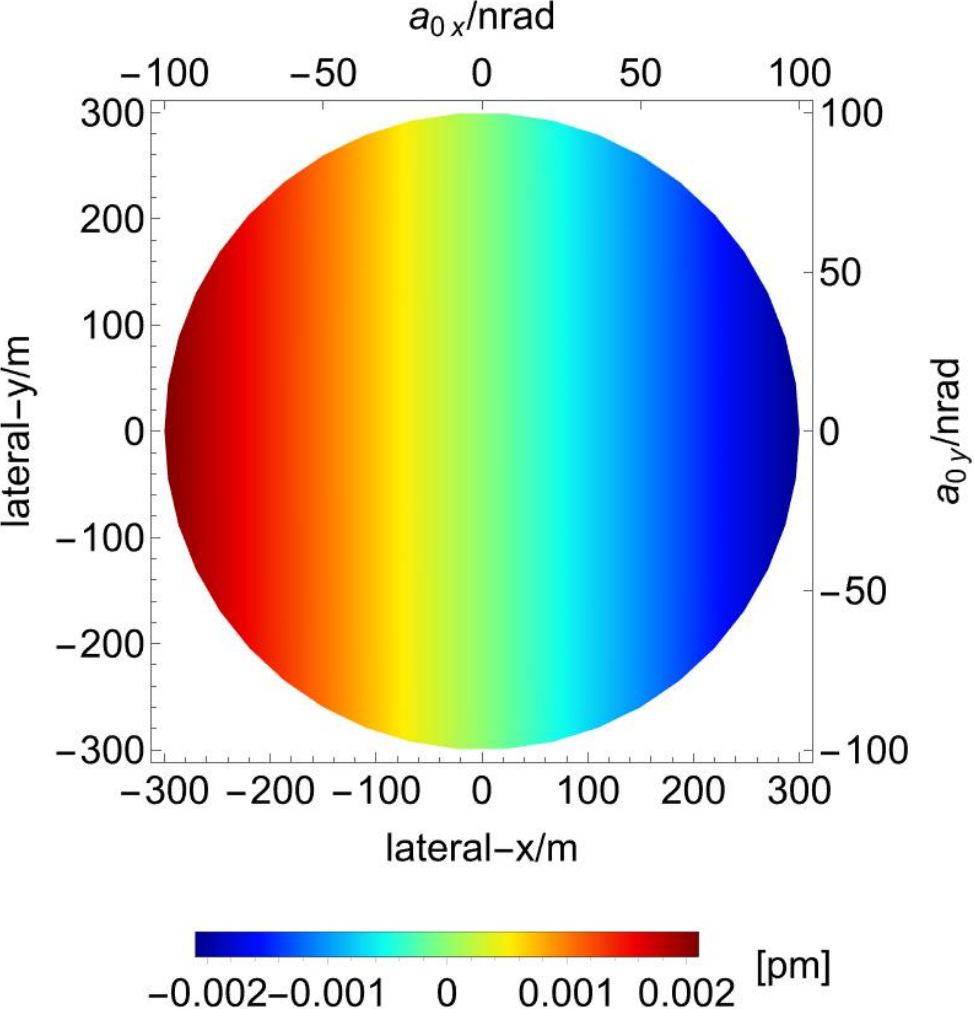}
			\label{fig:S1Z60}
		\end{subfigure}    
			 \begin{subfigure}[b]{0.3\textwidth}
			\centering
			\caption{$S^1Z_2^2$}
			\includegraphics[width=1\textwidth]{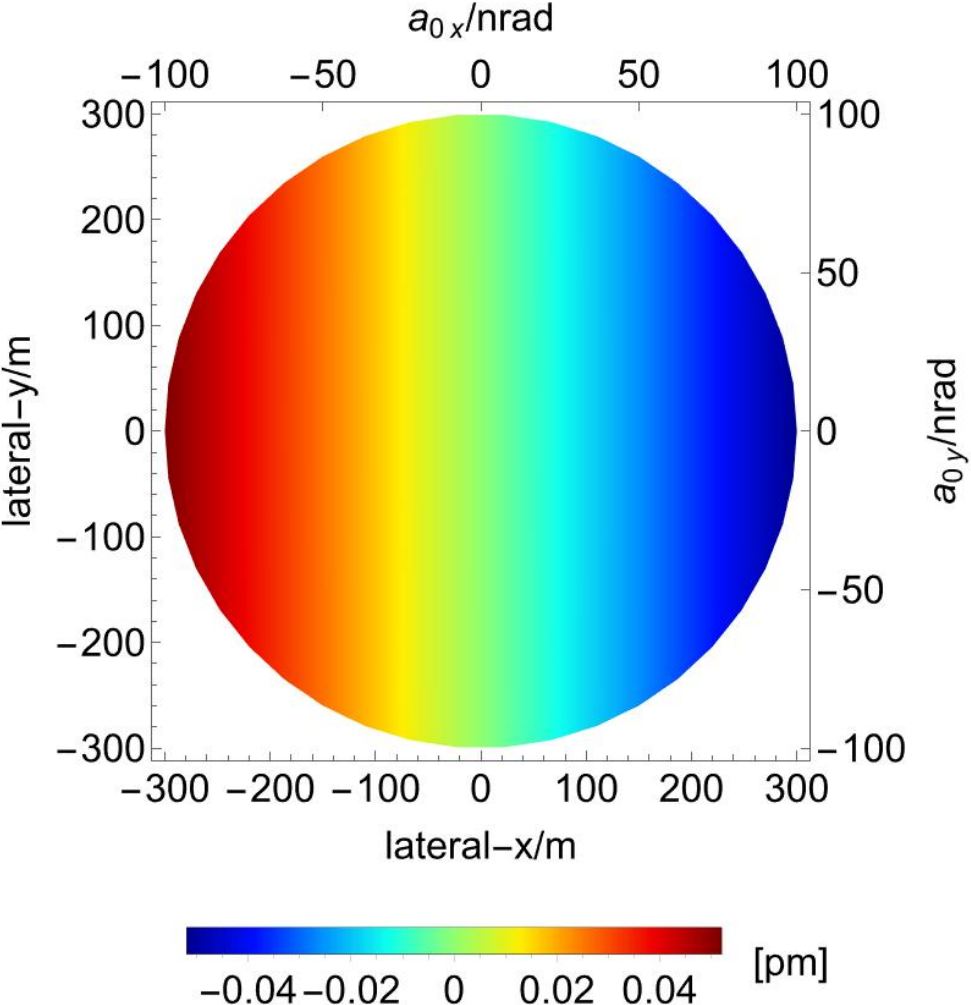}
			\label{fig:S1Z22}        
		\end{subfigure}
			\begin{subfigure}[b]{0.3\textwidth}
			\centering
			\caption{$S^1Z_4^2$}
			\includegraphics[width=1\textwidth]{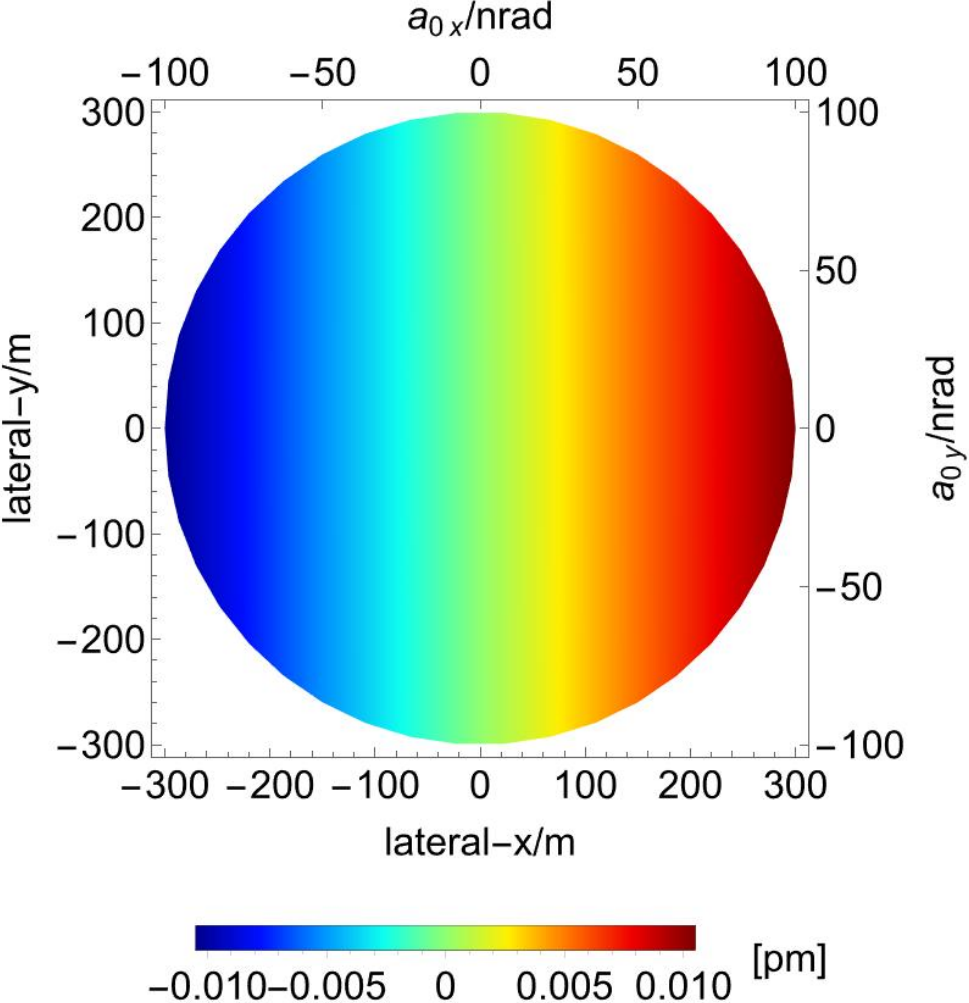}
			\label{fig:S1Z42}               
		\end{subfigure}
		 \caption{Far-field WFE induced by the second-order coupling between $S^1$ and the selected even-$n$ aberration terms. The transmitted WFE of each aberration is constrained to $\lambda/10$, with $a_4^0=0.418879$ and the other coefficients set to 0.314159. For the terms $Z_2^{\pm2}$ and $Z_4^{\pm2}$, only $Z_2^{2}$ and $Z_4^{2}$ are displayed.}
		 \label{fig:2order_WFEeven}
\end{figure}

We then calculated the contributions of the second-order coupling terms in \eqref{SZn0}--\eqref{SZn1} to the far-field WFE, as shown in Fig. \ref{fig:2order_WFEodd} and Fig. \ref{fig:2order_WFEeven}. The baseline case was set to $s_r=0.001$, and the initial level of each aberration term was constrained to $\lambda/10$. Because the couplings with even-$m$ aberrations act only on the real part of the expression, $Z_{2}^0$ was used as an auxiliary reference for evaluation. The initial aberration level of $Z_{2}^0$ was also set to $\lambda/10$.

The results show that the far-field WFE induced by the coupling terms between $S^{\pm1}$ and the aberration terms remains below the level that would affect the measurement. Even when $s_r$ is increased to 0.01, these contributions reach only the level of correction terms. Therefore, coupling terms higher than first order can be neglected.

\subsection{Influence on far-field TTL coupling noise} \label{sbse:4.3}
\begin{figure}
		\centering	   
		\begin{subfigure}[b]{0.35\textwidth}
			\centering
			\caption{Without beam shift}
			\includegraphics[width=1\textwidth]{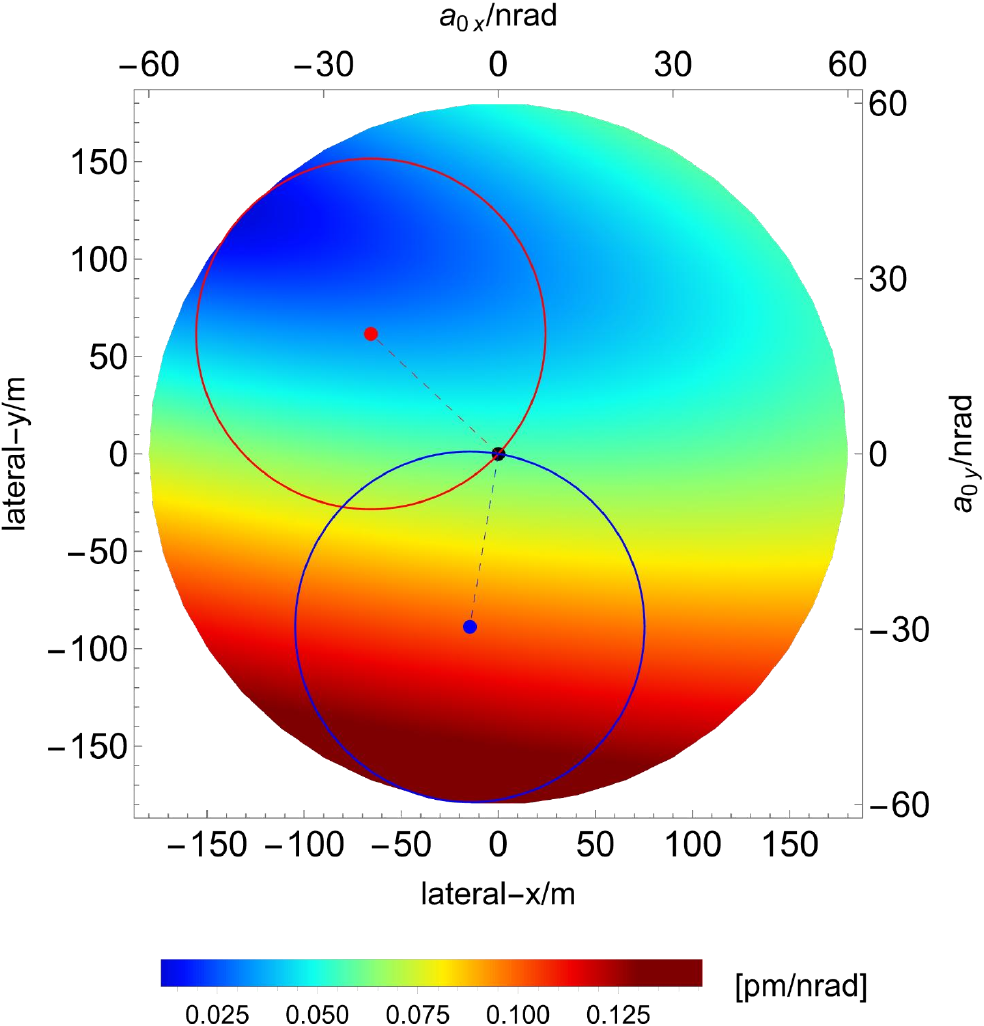}
			\label{fig:noise_level_o}
		\end{subfigure}
		\par\medskip
		\begin{subfigure}[b]{0.35\textwidth}
			\centering
			\caption{$s_r=0.001;{\theta}_0=\pi/2$}
			\includegraphics[width=1\textwidth]{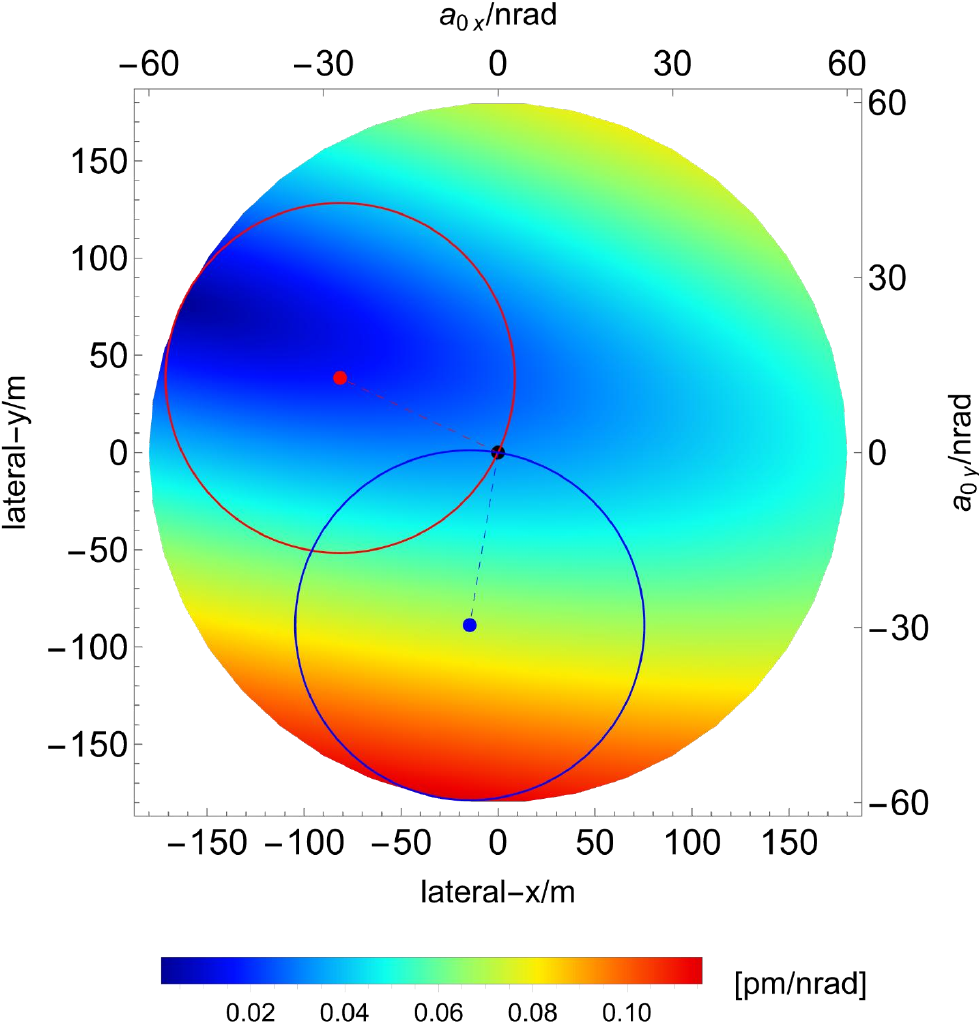}
			\label{fig:noise_level_l}
		\end{subfigure}
		\hspace{0.01\textwidth}
		\begin{subfigure}[b]{0.35\textwidth}
			\centering
			\caption{$s_r=0.001;{\theta}_0=-\pi/2$}
			\includegraphics[width=1\textwidth]{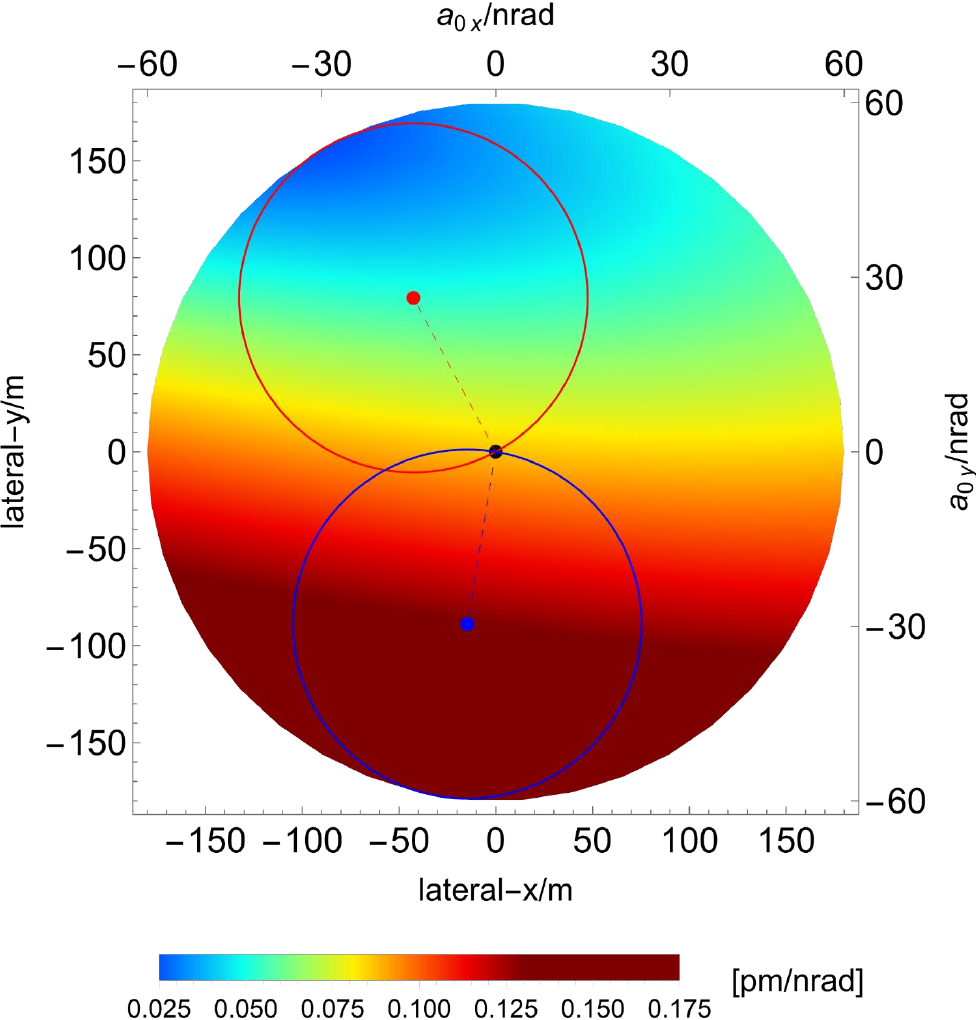}
			\label{fig:noise_level_h}
		\end{subfigure}   
		 \caption{Distribution of the phase-angle coupling coefficient for the far-field TTL coupling noise. The calculation adopts the Taiji arm length and telescope aperture with $q=0.9$, and the initial WFE is constructed from the Zernike coefficients listed in Table~\ref{Exampledcoefficients1}. The three panels correspond to the cases without beam shift and with $s_r=0.001$ along ${\theta}_0=\pm\pi/2$. Both the laser jitter and the static pointing offset levels are conservatively set to $30~\mathrm{nrad}/\sqrt{\mathrm{Hz}}$; therefore, the distribution is evaluated within a total fluctuation range of $60~\mathrm{nrad}/\sqrt{\mathrm{Hz}}$.}
		 \label{fig:noise_level}
	\end{figure}

The coupling noise level is calculated by incorporating the distortion terms $U_1(r, \psi, z)$. Subplot~\ref{fig:noise_level_o} in Fig.~\ref{fig:noise_level} shows the coupling noise level associated with the initial WFE constructed from the aberration coefficients listed in Table~\ref{Exampledcoefficients1}. This noise level is characterized by the phase-angle coupling coefficient, which is defined as follows:
\begin{equation}\label{NoiseLevel}
	{\delta}({\theta}_x,{\theta}_y)= \Vert \nabla ({W}_E({\theta}_x,{\theta}_y)) \Vert =\sqrt{ {(\frac{\partial{{W}_E({\theta}_x,{\theta}_y)}}{\partial{{\theta}_x}})}^2+{(\frac{\partial{{W}_E({\theta}_x,{\theta}_y)}}{\partial{{\theta}_y}})}^2}.
	\end{equation}

	In this example, the Taiji arm length and telescope aperture are again adopted, with $q=0.9$. Two radial beam shifts along the $\pm y$ directions are then introduced on the basis of the distortion term, corresponding to subplots~\ref{fig:noise_level_l} and \ref{fig:noise_level_h} in Fig.~\ref{fig:noise_level}, where $s_r$ is set to $0.001$. As shown in subplots~\ref{fig:noise_level_l} and \ref{fig:noise_level_h}, the beam-shift term may either increase or decrease the overall noise level. This occurs because the aberrations induced by the beam shift are generally uncorrelated with those induced by the distortion term. As a result, the superposition or cancellation of these aberrations depends on the actual beam-shift direction and the orientation of the asymmetric aberration terms. Since these factors are difficult to control in practice, the worst-case condition should be considered in the performance analysis.
\begin{table*}
	\sisetup{round-mode=places, round-precision=5}
	\abovetopsep=0pt
	\aboverulesep=0pt
	\belowrulesep=0pt
	\belowbottomsep=0pt
	\renewcommand\arraystretch{1.5}
		\begin{center}
			\setlength{\tabcolsep}{1mm}
			\begin{tabular}{c|c c c c c c c}
				\toprule[1.5pt]
				${Z^m_n}$ &  ${Z_1^1}$ &  ${Z_1^{-1}}$ & ${Z_2^{0}}$ &  ${Z_2^2}$ &  ${Z_2^{-2}}$ &  ${Z_3^1}$ &  ${Z_3^{-1}}$\\ 
				\midrule[1.5pt]
				${a^m_n}$ & $\num{-0.0290403}$ & $\num{0.0135543}$ & $\num{0.145519}$ & $\num{-0.105273}$& $\num{0.0566279}$ & $\num{0.0158143}$ & $\num{-0.048233}$ \\
				\midrule[1.5pt]
				 &  ${Z_3^3}$ &  ${Z_3^{-3}}$ & ${Z_4^{0}}$ &  ${Z_4^2}$ &  ${Z_4^{-2}}$ &  ${Z_4^4}$ &  ${Z_4^{-4}}$ \\ 
				\midrule[1.5pt]
				 & $\num{-0.112372}$ & $\num{-0.0447792}$ & $\num{-0.0677458}$ & $\num{0.0158143}$& $\num{0.0124993}$ & $\num{0.0975962}$ & $\num{0.00259626}$ \\
				\midrule[1.5pt]
				 &  ${Z_5^1}$ &  ${Z_5^{-1}}$ & ${Z_5^{3}}$ &  ${Z_5^{-3}}$ &  ${Z_5^{5}}$ &  ${Z_5^{-5}}$ &  ${Z_6^{0}}$\\ 
				\midrule[1.5pt]
				 & $\num{0.0689143}$ & $\num{-0.0733618}$ & $\num{0.0318497}$ & $\num{-0.0299652}$& $\num{-0.0109276}$ & $\num{0.00481528}$ & $\num{0.0158143}$ \\
				\bottomrule[1.5pt]
			\end{tabular}
			\caption{Zernike aberration coefficients used in the calculation. The corresponding distribution of the phase-angle coupling coefficient is shown in Fig.~\ref{fig:noise_level_o}.}
	\label{Exampledcoefficients1}
		\end{center}
	\end{table*}

\section{Conclusions} \label{se:5}

In this paper, we extended the Nijboer--Zernike analytical framework for evaluating far-field wavefront error (WFE) in spaceborne laser interferometric links by introducing two practical initial-condition parameters: the beam-waist-to-aperture ratio $q$ and the normalized lateral spot-shift ratio $s_r$. These parameters describe, respectively, the truncation condition of the emitted Gaussian beam and the lateral displacement between the beam spot center and the telescope pupil center. The extended model provides a quantitative way to evaluate how beam-parameter selection and alignment errors affect far-field WFE and the associated TTL coupling.

For the beam-waist-to-aperture ratio $q$, the received on-axis optical power reaches its maximum at $q=0.892135$, supporting the commonly adopted choice of $q\approx0.9$. This power-based criterion remains the dominant consideration in spaceborne laser interferometric links, since weak received light directly limits the subsequent phase-locking and interferometric readout performance. Although our analysis shows that decreasing $q$ within $0.8\le q\le1$ can slightly reduce the far-field WFE caused by random initial aberrations, the reduction is moderate compared with the importance of maintaining sufficient received optical power. Therefore, the present result should be understood as a quantitative supplement to the conventional power-based choice of $q$, rather than as a replacement of it. Overall, $q\approx0.9$ remains an appropriate and balanced design choice.

For the normalized lateral spot-shift ratio $s_r$, the direct spot-shift term is found to be the dominant contribution. Even in the absence of transmitted WFE, a lateral displacement of the beam spot introduces an additional far-field WFE and therefore contributes directly to TTL coupling. For $q=0.892135$, when $s_r=0.001$, the corresponding phase-angle coupling coefficient is approximately $0.0892~\mathrm{pm/nrad}$, which is already close to the typical far-field TTL coupling requirement of $0.1~\mathrm{pm/nrad}$. Taking a Taiji-like telescope as an example, this value corresponds to a beam displacement of only about $2~\mu\mathrm{m}$ at the entrance pupil. This indicates that the lateral centering of the transmitted beam can impose a stringent alignment requirement and should be treated as an important tolerance item in optical design and assembly.

The coupling between the lateral spot shift and the transmitted WFE was also derived and estimated for the dominant Zernike aberration terms. The results show that these coupling contributions remain much smaller than the direct spot-shift contribution under the considered conditions. Even for a larger shift ratio, their effect stays at the level of correction terms and does not significantly change the far-field WFE. After including the direct spot-shift term, the overall far-field TTL phase-angle coupling coefficient increases, indicating that this term can significantly raise the total noise level. Nevertheless, since the contribution introduced by the direct spot shift is linear, it is expected to be easier to calibrate and subtract through data post-processing than nonlinear coupling effects. These conclusions provide guidance for related experimental design, optical assembly, and future space-based gravitational-wave detection missions.

\section*{Acknowledgements.}
This work has been supported in part by the National Key Research and Development Program of China under Grant No. 2020YFC2201501, the National Science Foundation of China (NSFC) under Grants No. 12147103 (special fund to the center for quanta-to-cosmos theoretical physics), No. 11821505, the Strategic Priority Research Program of the Chinese Academy of Sciences under Grant No. XDB23030100, and the Chinese Academy of Sciences (CAS).
\appendix
\section{Expressions and coefficients of the first 21 aberration terms and their couplings with the $q$-correction} \label{se:6}
We denote $(-1)^l(2l+1)I_{l+1/2}(\frac{1}{2q^2})$ by $L_l$.
\begin{itemize}
\item[$\blacksquare$]${Z^{2\beta}_{{\gamma}{'}}}$
\begin{equation}\label{Zn0}
\begin{aligned}
Z_n^{0}(r, \psi, z)=qe^{-\frac{1}{2q^2}}{\pi}^{\frac{1}{2}}(2\pi)\left\{
i\left[{\alpha}_n^{0}\frac{J_1(v)}{v}+{\beta}_n^{0}\frac{J_3(v)}{v}\right]
+\delta_{n,2}\left[{\bar{\alpha}}_2^{0}\frac{J_1(v)}{v}+{\bar{\beta}}_2^{0}\frac{J_3(v)}{v}\right]\right\},
\end{aligned}
\end{equation}
\begin{equation}\label{Zn2}
\begin{aligned}
	\begin{pmatrix}
			Z_n^{2} \\
			Z_n^{-2} \\
		\end{pmatrix}
(r, \psi, z)=qe^{-\frac{1}{2q^2}}{\pi}^{\frac{1}{2}}(2\pi)\left\{-i
		\begin{pmatrix}
			\cos2\psi \\
			\sin2\psi \\
		\end{pmatrix}
		\left[{\alpha}_n^{\pm2}\frac{J_1(v)}{v}+{\beta}_n^{\pm2}\frac{J_3(v)}{v}\right]\right\},
\end{aligned}
\end{equation}
where
\begin{gather*}\label{coEven}
\left({\alpha}_2^0,\;{\beta}_2^0\right)=a_2^0\left(0.333333L_1,-L_0-0.4L_2\right)-\frac{(a_2^0)^3}{6}\left(0.2L_1+0.057143L_3,\;-0.6L_0-0.342857L_2\right),\\
\left({\bar{\alpha}}_2^0,\;{\bar{\beta}}_2^0\right)=-\frac{(a_2^0)^2}{2}\left(0.333333L_0+0.133333L_2,\;-0.6L_1-0.171429L_3\right),\\
\left({\alpha}_2^{\pm2},\;{\beta}_2^{\pm2}\right)={a_2^{\pm2}}\left(0,\;L_0+0.5L_1+0.1L_2\right),\\
\left({\alpha}_4^{0},\;{\beta}_4^{0}\right)={a_4^0}\left(0.2L_2,\;-0.4L_1-0.257143L_3\right),\\
\left({\alpha}_4^{\pm2},\;{\beta}_4^{\pm2}\right)={a_4^{\pm2}}\left(0,\;0.3L_1+0.3L_2+0.0857143L_3\right),\\
\left({\alpha}_6^{0},\;{\beta}_6^{0}\right)=a_6^0\left(0.142857L_3,\;-0.257143L_2\right).
\end{gather*}
Here, the quadratic real-part term $\left({\bar{\alpha}}_2^0,\;{\bar{\beta}}_2^0\right)$ is retained only for $Z_2^0$; for the other $Z_n^0$ terms, only the imaginary contribution is retained at this order.
\end{itemize}

\begin{itemize}
\item[$\blacksquare$]${Z^{2\alpha+1}_{\gamma}}$

Similarly, for ${Z^{2\alpha+1}_{\gamma}}$, we obtain:
\begin{equation}\label{Zn1}
\begin{aligned}
	\begin{pmatrix}
			Z_n^{1} \\
			Z_n^{-1} \\
		\end{pmatrix}
(r, \psi, z)=qe^{-\frac{1}{2q^2}}{\pi}^{\frac{1}{2}}(2\pi)\left\{
		\begin{pmatrix}
			\cos\psi \\
			\sin\psi \\
		\end{pmatrix}
		\left[{\sigma}_n^{\pm1}\frac{J_2(v)}{v}+{\tau}_n^{\pm1}\frac{J_4(v)}{v}\right]\right\},
\end{aligned}
\end{equation}
\begin{equation}\label{Z33}
\begin{aligned}
	\begin{pmatrix}
			Z_n^{3} \\
			Z_n^{-3} \\
		\end{pmatrix}
(r, \psi, z)=qe^{-\frac{1}{2q^2}}{\pi}^{\frac{1}{2}}(2\pi)\left\{-
	\begin{pmatrix}
			\cos3\psi \\
			\sin3\psi \\
		\end{pmatrix}
\left[{\sigma}_n^{\pm3}\frac{J_2(v)}{v}+{\tau}_n^{\pm3}\frac{J_4(v)}{v}\right]\right\},
\end{aligned}
\end{equation}
where
\begin{gather*}\label{coEven_appendix}
\left({\sigma}_1^{\pm1},\;{\tau}_1^{\pm1}\right)=a_1^{\pm1}\left(L_0+0.333333L_1,\;-0.666667L_1-0.4L_2\right),\\
\left({\sigma}_3^{\pm1},\;{\tau}_3^{\pm1}\right)=a_3^{\pm1}\left(0.333333L_1+0.2L_2,\;-L_0-0.0666667L_1-0.2L_2-0.257143L_3\right),\\
\left({\sigma}_3^{\pm3},\;{\tau}_3^{\pm3}\right)=a_3^{\pm3}\left(0,\;L_0+0.6L_1+0.2L_2+0.028571L_3\right),\\
\left({\sigma}_5^{\pm1},\;{\tau}_5^{\pm1}\right)=a_5^{\pm1}\left(0.2L_2+0.142857L_3,\;-0.4L_1-0.057143L_2-0.114286L_3\right),\\
\left({\sigma}_5^{\pm3},\;{\tau}_5^{\pm3}\right)=a_5^{\pm3}\left(0,\;0.266667L_1+0.342857L_2+0.171429L_3\right).
\end{gather*}
\end{itemize}

\begin{itemize}
\item[$\blacksquare$]${Z^{2\alpha+1}_{\gamma}}{Z^{2\beta}_{{\gamma}{'}}}$

For ${Z^{2\alpha+1}_{\gamma}}{Z^{2\beta}_{{\gamma}{'}}}$, we obtain:
\begin{equation}\label{Zn1Zn'0}
	\begin{pmatrix}
			Z_n^{1}Z_{n{'}}^{0} \\
			Z_n^{-1}Z_{n{'}}^{0} \\
		\end{pmatrix}
		(r, \psi, z)=qe^{-\frac{1}{2q^2}}{\pi}^{\frac{1}{2}}(2\pi)\left\{i
	\begin{pmatrix}
			\cos\psi \\
			\sin\psi \\
		\end{pmatrix}
	\left[{\alpha}_{n;n{'}}^{\pm1;0}\frac{J_2(v)}{v}+{\beta}_{n;n{'}}^{\pm1;0}\frac{J_4(v)}{v}\right]\right\},
\end{equation}
\begin{equation}\label{Zn1Zn'pm2}
\begin{aligned}
	&\begin{pmatrix}
			Z_n^{1}Z_{n{'}}^{2} &  Z_n^{1}Z_{n{'}}^{-2}\\
			Z_n^{-1}Z_{n{'}}^{2} & Z_n^{-1}Z_{n{'}}^{-2}\\
		\end{pmatrix}
		(r, \psi, z)=qe^{-\frac{1}{2q^2}}{\pi}^{\frac{1}{2}}(2\pi)\\
		&\left\{\frac{i}{2}
	\begin{pmatrix}
			\cos\psi &  \sin\psi\\
			-\sin\psi & \cos\psi\\
		\end{pmatrix}
			\left[{{\alpha}_1}_{n;n{'}}^{\pm1;\pm2}\frac{J_2(v)}{v}+{{\beta}_1}_{n;n{'}}^{\pm1;\pm2}\frac{J_4(v)}{v}\right]-\frac{i}{2}
		\begin{pmatrix}
			\cos3\psi &  \sin3\psi\\
			\sin3\psi & -\cos3\psi\\
		\end{pmatrix}				
			\left[{{\alpha}_2}_{n;n{'}}^{\pm1;\pm2}\frac{J_2(v)}{v}+{{\beta}_2}_{n;n{'}}^{\pm1;\pm2}\frac{J_4(v)}{v}\right]\right\},	
\end{aligned}
\end{equation}
\begin{equation}\label{Zn1Zn'pm4}
\begin{aligned}
	&\begin{pmatrix}
			Z_n^{1}Z_{n{'}}^{4} &  Z_n^{1}Z_{n{'}}^{-4}\\
			Z_n^{-1}Z_{n{'}}^{4} & Z_n^{-1}Z_{n{'}}^{-4}\\
		\end{pmatrix}
		(r, \psi, z)=qe^{-\frac{1}{2q^2}}{\pi}^{\frac{1}{2}}(2\pi)
		\left\{-\frac{i}{2}
		\begin{pmatrix}
			\cos3\psi &  \sin3\psi\\
			-\sin3\psi & \cos3\psi\\
		\end{pmatrix}				
			\left[{{\alpha}}_{n;n{'}}^{\pm1;\pm4}\frac{J_2(v)}{v}+{{\beta}}_{n;n{'}}^{\pm1;\pm4}\frac{J_4(v)}{v}\right]\right\},	
\end{aligned}
\end{equation}
\begin{equation}\label{Zn3Zn'0}
	\begin{pmatrix}
			Z_n^{3}Z_{n{'}}^{0} \\
			Z_n^{-3}Z_{n{'}}^{0} \\
		\end{pmatrix}
		(r, \psi, z)=qe^{-\frac{1}{2q^2}}{\pi}^{\frac{1}{2}}(2\pi)\left\{-i
	\begin{pmatrix}
			\cos3\psi \\
			\sin3\psi \\
		\end{pmatrix}
	\left[{\alpha}_{n;n{'}}^{\pm3;0}\frac{J_2(v)}{v}+{\beta}_{n;n{'}}^{\pm3;0}\frac{J_4(v)}{v}\right]\right\},	
\end{equation}
\begin{equation}\label{Zn3Zn'pm2}
\begin{aligned}
	&\begin{pmatrix}
			Z_n^{3}Z_{n{'}}^{2} &  Z_n^{3}Z_{n{'}}^{-2}\\
			Z_n^{-3}Z_{n{'}}^{2} & Z_n^{-3}Z_{n{'}}^{-2}\\
		\end{pmatrix}
		(r, \psi, z)=qe^{-\frac{1}{2q^2}}{\pi}^{\frac{1}{2}}(2\pi)
		\left\{\frac{i}{2}
	\begin{pmatrix}
			\cos\psi &  -\sin\psi\\
			\sin\psi & \cos\psi\\
		\end{pmatrix}
			\left[{{\alpha}}_{n;n{'}}^{\pm3;\pm2}\frac{J_2(v)}{v}+{{\beta}}_{n;n{'}}^{\pm3;\pm2}\frac{J_4(v)}{v}\right]\right\},	
\end{aligned}
\end{equation}
\begin{equation}\label{Zn3Zn'pm4}
\begin{aligned}
	&\begin{pmatrix}
			Z_n^{3}Z_{n{'}}^{4} &  Z_n^{3}Z_{n{'}}^{-4}\\
			Z_n^{-3}Z_{n{'}}^{4} & Z_n^{-3}Z_{n{'}}^{-4}\\
		\end{pmatrix}
		(r, \psi, z)=qe^{-\frac{1}{2q^2}}{\pi}^{\frac{1}{2}}(2\pi)
		\left\{\frac{i}{2}
		\begin{pmatrix}
			\cos\psi &  \sin\psi\\
			-\sin\psi & \cos\psi\\
		\end{pmatrix}				
			\left[{{\alpha}}_{n;n{'}}^{\pm3;\pm4}\frac{J_2(v)}{v}+{{\beta}}_{n;n{'}}^{\pm3;\pm4}\frac{J_4(v)}{v}\right]\right\},	
\end{aligned}
\end{equation}
\begin{equation}\label{Zn5Zn'pm2}
\begin{aligned}
	&\begin{pmatrix}
			Z_n^{5}Z_{n{'}}^{2} &  Z_n^{5}Z_{n{'}}^{-2}\\
			Z_n^{-5}Z_{n{'}}^{2} & Z_n^{-5}Z_{n{'}}^{-2}\\
		\end{pmatrix}
		(r, \psi, z)=qe^{-\frac{1}{2q^2}}{\pi}^{\frac{1}{2}}(2\pi)
		\left\{-\frac{i}{2}
	\begin{pmatrix}
			\cos3\psi &  -\sin3\psi\\
			\sin3\psi & \cos3\psi\\
		\end{pmatrix}
			\left[{{\alpha}}_{n;n{'}}^{\pm5;\pm2}\frac{J_2(v)}{v}+{{\beta}}_{n;n{'}}^{\pm5;\pm2}\frac{J_4(v)}{v}\right]\right\},	
\end{aligned}
\end{equation}
\begin{equation}\label{Zn5Zn'pm4}
\begin{aligned}
	&\begin{pmatrix}
			Z_n^{5}Z_{n{'}}^{4} &  Z_n^{5}Z_{n{'}}^{-4}\\
			Z_n^{-5}Z_{n{'}}^{4} & Z_n^{-5}Z_{n{'}}^{-4}\\
		\end{pmatrix}
		(r, \psi, z)=qe^{-\frac{1}{2q^2}}{\pi}^{\frac{1}{2}}(2\pi)
		\left\{\frac{i}{2}
	\begin{pmatrix}
			\cos\psi &  -\sin\psi\\
			\sin\psi & \cos\psi\\
		\end{pmatrix}
			\left[{{\alpha}}_{n;n{'}}^{\pm5;\pm4}\frac{J_2(v)}{v}+{{\beta}}_{n;n{'}}^{\pm5;\pm4}\frac{J_4(v)}{v}\right]\right\}.	
\end{aligned}
\end{equation}
The corresponding coefficients are listed in Table \ref{2ndcoefficients}.
\begin{sidewaystable*}
\tiny
\abovetopsep=0pt
\aboverulesep=0pt
\belowrulesep=0pt
\belowbottomsep=0pt
\renewcommand\arraystretch{2.5}
	\begin{center}
			\setlength{\tabcolsep}{0.7mm}
		\begin{tabular}[width=0.7\textwidth]{c|c c c}
			\midrule[1pt]
			 & $Z_2^{0}$ &  $Z_2^{\pm2}$  \\ 
			 \midrule[1pt]
			$Z_1^{\pm1}$ & $(0.333333L_0+0.333333L_1+0.133333L_2,\;-0.666667L_0-0.666667L_1-0.666667L_2-0.171429L_3)$ & \makecell[c]{${(0.666667L_0+0.333333L_1+0.666667L_2,\;-0.333333L_0-0.466667L_1-0.333333L_2-0.085714L_3)}_1$\\${(0.,\;L_0+0.6L_1+0.2L_2+0.028571L_3)}_2$}\\
			\hline
			$Z_3^{\pm1}$ & $(0.333333L_0+0.133333L_1+0.133333L_2+0.085714L_3,\;-0.066667L_0-0.466667L_1-0.180952L_2-0.085714L_3)$ & \makecell[c]{${(0.166667L_0+0.233333L_1+0.166667L_2+0.042857L_3,\;-0.533333L_0-0.266667L_1-0.190476L_2-0.171429L_3)}_1$\\${(0.,\;0.4L_0+0.4L_1+0.285714L_2+0.114286L_3)}_2$} \\
			\hline
			$Z_3^{\pm3}$ &  $(0,\;0.6L_0+0.466667L_1+0.257143L_2+0.085714L_3)$ &  $(0.5L_0+0.3L_1+0.1L_2+0.014286L_3,\;-0.4L_0-0.4L_1-0.285714L_2-0.114286L_3)$\\
			\hline
			$Z_5^{\pm1}$  & $(0.1333334L_1+0.085714L_2+0.085714L_3,\;-0.4L_0-0.038095L_1-0.228571L_2-0.133333L_3)$ & \makecell[c]{${(0.066667L_1+0.142857L_2+0.114286L_3,\;-0.2L_0-0.219048L_1-0.142857L_2-0.123810L_3)}_1$\\${(0.,\;0.066667L_0+0.161905L_1+0.238095L_2+0.193651L_3)}_2$}\\
			\hline
			$Z_5^{\pm3}$ &  $(0,\;0.266667L_0+0.228571L_1+0.209524L_2+0.165079L_3)$ &  $(0.133333L_1+0.171429L_2+0.085714L_3,\;-0.4L_0-0.209524L_1-0.142857L_2-0.161905L_3)$\\
			\hline
			$Z_5^{\pm5}$ &  $\backslash$ & $(0,\;0.666667L_0+0.476190L_1+0.238095L_2+0.079365L_3)$\\
			\midrule[1pt]
			\midrule[1pt]
			&  $Z_4^0$ &  $Z_4^{\pm2}$\\
			\midrule[1pt]
			$Z_1^{\pm1}$ &  $(0.133333L_1+0.2L_2+0.085714L_3,\;-0.4L_0-0.266667L_1-0.114286L_2-0.171429L_3)$ & \makecell[c]{${(0.2L_1+0.2L_2+0.057143L_3,\;-0.6L_0-0.2L_1-0.142857L_2-0.2L_3)}_1$\\${(0.,\;0.2L_0+0.333333L_1+0.31428L_2+0.142857L_3)}_2$}\\
			\hline
			$Z_3^{\pm1}$  &  $(0.2L_0+0.133333L_1+0.057143L_2+0.085714L_3,\;-0.2L_0-0.180952L_1-0.257143L_2-0.085714L_3)$ & \makecell[c]{${(0.3L_0+0.1L_1+0.071429L_2+0.1L_3,\;-0.342857L_1-0.228571L_2-0.057143L_3)}_1$\\${(0.,\;0.4L_0+0.247619L_1+0.171429L_2+0.152381L_3)}_2$}\\
			\hline
			$Z_3^{\pm3}$  &  $(0,\;0.2L_0+0.257143L_1+0.257143L_2+0.16190L_3)$ &  $(0.1L_0+0.166667L_1+0.157143L_2+0.071429L_3,\;-0.4L_0-0.247619L_1-0.171429L_2-0.152381L_3)$\\
			\hline
			$Z_5^{\pm1}$  &  $(0.2L_0+0.085714L_1+0.057143L_2+0.038095L_3,\;-0.057143L_0-0.228571L_1-0.114286L_2-0.133333L_3)$ & \makecell[c]{${(0.133333L_0+0.123810L_1+0.047619L_2+0.044444L_3,\;-0.238095L_0-0.104762L_1-0.180952L_2-0.136508L_3)}_1$\\${(0.,\;0.257143L_0+0.2L_1+0.123810L_2+0.104762L_3)}_2$}\\
			\hline
			$Z_5^{\pm3}$  &  $(0,\;0.342857L_0+0.209524L_1+0.114286L_2+0.114286L_3)$ &  $(0.266667L_0+0.095238L_1+0.038095L_2+0.079365L_3,\;-0.019048L_0-0.247619L_1-0.219048L_2-0.073016L_3)$\\
			\hline
			$Z_5^{\pm5}$ &  $\backslash$ & $(0,\;0.285714L_0+0.285714L_1+0.238095L_2+0.142857L_3)$\\
			\midrule[1pt]
			\midrule[1pt]
			&  $Z_4^{\pm4}$ &  $Z_6^0$\\
			\midrule[1pt]
			$Z_1^{\pm1}$ &  $(0.,\;0.8L_0+0.533333L_1+0.228571L_2+0.057143L_3)$ & $(0.085714L_2+0.142857L_3,\;-0.171429L_1-0.171429L_2-0.076191L_3)$\\
			\hline
			$Z_3^{\pm1}$  &  $(0.,\;0.4L_0+0.361905L_1+0.257143L_2+0.123810L_3)$ & $(0.085714L_1+0.085714L_2+0.038095L_3,\;-0.257143L_0-0.085714L_1-0.085714L_2-0.180952L_3)$\\
			\hline
			$Z_3^{\pm3}$  & $(0.4L_0+0.266667L_1+0.114286L_2+0.028571L_3,\;-0.4L_0-0.361905L_1-0.257143L_2-0.123810L_3)$ & $(0.,\;0.028571L_0+0.085714L_1+0.161905L_2+0.180952L_3)$\\
			\hline
			$Z_5^{\pm1}$  &  $(0.,\;0.114286L_0+0.171429L_1+0.209524L_2+0.171429L_3)$ & $(0.142857L_0+0.085714L_1+0.038095L_2+0.038095L_3,\;-0.114286L_0-0.133333L_1-0.133333L_2-0.055411L_3)$\\
			\hline
			$Z_5^{\pm3}$  &  $(0.066667L_0+0.123810L_1+0.138095L_2+0.084127L_3,\;-0.304762L_0-0.219048L_1-0.161905L_2-0.139683L_3)$ & $(0.,\;0.171429L_0+0.165079L_1+0.114286L_2+0.072727L_3)$\\
			\hline
			$Z_5^{\pm5}$ & $(0.333333L_0+0.238095L_1+0.119048L_2+0.039683L_3,\;-0.380952L_0-0.333333L_1-0.238095L_2-0.126984L_3)$ & $\backslash$\\
			\midrule[1pt]
		\end{tabular}
		\caption{The coefficient list of $Z^{2\alpha+1}_{\gamma}Z^{2\beta}_{{\gamma}{'}}$. Each term should be multiplied by the corresponding Zernike coefficients $a^{2\alpha+1}_{\gamma}$ and $a^{2\beta}_{{\gamma}{'}}$. "${(\quad)}_1$" and "${(\quad)}_2$" correspond to the coefficients $({{\alpha}_1}_{n;n{'}}^{\pm1;\pm2},{{\beta}_1}_{n;n{'}}^{\pm1;\pm2})$ and $({{\alpha}_2}_{n;n{'}}^{\pm1;\pm2},{{\beta}_2}_{n;n{'}}^{\pm1;\pm2})$ in \eqref{Zn1Zn'pm2}.}
\label{2ndcoefficients}
	\end{center}
\end{sidewaystable*}
\end{itemize}

\begin{itemize}
\item[$\blacksquare$]${Z^{2{\alpha}_1+1}_{{\gamma}_1}}{Z^{2{\alpha}_2+1}_{{\gamma}_2}}$

For ${Z^{2{\alpha}_1+1}_{{\gamma}_1}}{Z^{2{\alpha}_2+1}_{{\gamma}_2}}$, we only keep
$Z_1^{\pm1}Z_1^{\pm1}$, $Z_1^{\pm1}Z_3^{\pm1}$, $Z_1^{\pm1}Z_5^{\pm1}$, and $Z_3^{\pm1}Z_3^{\pm1}$:
\begin{equation}\label{Zn1Zn'1}
	\begin{aligned}
		&\begin{pmatrix}
				Z_n^{1}Z_{n{'}}^{1} \\
				Z_n^{-1}Z_{n{'}}^{-1} \\
			\end{pmatrix}
			(r, \psi, z)=e^{-\frac{1}{2}}{\pi}^{\frac{1}{2}}(2\pi)\\
			&\left\{-\frac{1}{2}\left[{{\alpha}_1}_{n;n{'}}^{\pm1;\pm1}\frac{J_1(v)}{v}+{{\beta}_1}_{n;n{'}}^{\pm1;\pm1}\frac{J_3(v)}{v}\right]+\frac{1}{2}
		\begin{pmatrix}
				\cos2\psi \\
				-\cos2\psi \\
			\end{pmatrix}
		\left[{{\alpha}_2}_{n;n{'}}^{\pm1;\pm1}\frac{J_1(v)}{v}+{{\beta}_2}_{n;n{'}}^{\pm1;\pm1}\frac{J_3(v)}{v}\right]\right\},
	\end{aligned}
	\end{equation}
where
\begin{gather*}\label{coOddOdd_appendix}
\left({{\alpha}_1}_{1;1}^{\pm1;\pm1},\;{{\beta}_1}_{1;1}^{\pm1;\pm1}\right)=a_1^{\pm1}a_1^{\pm1}\left(0.25L_0+0.0833335L_1,\;-0.25L_0-0.25L_1-0.1L_2\right),\\
\left({{\alpha}_2}_{1;1}^{\pm1;\pm1},\;{{\beta}_2}_{1;1}^{\pm1;\pm1}\right)=a_1^{\pm1}a_1^{\pm1}\left(0,\;0.5L_0+0.25L_1+0.05L_2\right),\\
\left({{\alpha}_1}_{1;3}^{\pm1;\pm1},\;{{\beta}_1}_{1;3}^{\pm1;\pm1}\right)=a_1^{\pm1}a_3^{\pm1}\left(0.166667L_1+0.1L_2,\;-0.5L_0-0.2L_1-0.2L_2-0.128571L_3\right),\\
\left({{\alpha}_2}_{1;3}^{\pm1;\pm1},\;{{\beta}_2}_{1;3}^{\pm1;\pm1}\right)=a_1^{\pm1}a_3^{\pm1}\left(0,\;0.25L_0+0.35L_1+0.25L_2+0.0642857L_3\right),\\
\left({{\alpha}_1}_{1;5}^{\pm1;\pm1},\;{{\beta}_1}_{1;5}^{\pm1;\pm1}\right)=a_1^{\pm1}a_5^{\pm1}\left(0.1L_2+0.0714286L_3,\;-0.2L_1-0.128571L_2-0.128571L_3\right),\\
\left({{\alpha}_2}_{1;5}^{\pm1;\pm1},\;{{\beta}_2}_{1;5}^{\pm1;\pm1}\right)=a_1^{\pm1}a_5^{\pm1}\left(0,\;0.1L_1+0.214286L_2+0.171429L_3\right),\\
\left({{\alpha}_1}_{3;3}^{\pm1;\pm1},\;{{\beta}_1}_{3;3}^{\pm1;\pm1}\right)=a_3^{\pm1}a_3^{\pm1}
\left(0.125L_0+0.00833335L_1+0.025L_2+0.0321429L_3,\right.\\
\left.\quad-0.025L_0-0.175L_1-0.067857L_2-0.0321429L_3\right),\\
\left({{\alpha}_2}_{3;3}^{\pm1;\pm1},\;{{\beta}_2}_{3;3}^{\pm1;\pm1}\right)=a_3^{\pm1}a_3^{\pm1}\left(0,\;0.2L_0+0.1L_1+0.0714285L_2+0.0642855L_3\right).
\end{gather*}
\end{itemize}

\bibliography{reference}
\makeatletter
\end{document}